\documentclass[
journal,
comsoc
]{IEEEtran}
\usepackage{tabularx}
\usepackage[normalem]{ulem}
\usepackage{cite}
\usepackage{amsmath,amssymb,amsfonts}
\usepackage{algorithmic}
\usepackage[usenames]{color}
\usepackage{psfrag}
\usepackage{url}
\usepackage{comment}
\usepackage{graphicx,color}
 \usepackage{epstopdf}
 \usepackage[nolist]{acronym}
 \usepackage{graphicx}
 \usepackage{float} 
 \usepackage{subfig}
 \usepackage{booktabs}
 \usepackage{multirow}
\usepackage[capitalise]{cleveref}
\Crefname{equation}{Eq.\!}{Eqs.\!}
\Crefname{figure}{Fig.\!}{Figs.\!}
\Crefname{tabular}{Tab.\!}{Tabs.\!}
\Crefname{section}{Section\!}{Sections.\!}
\usepackage{pgfplots}
\usepackage{pgfplotstable}
\usepgfplotslibrary{patchplots}
\usetikzlibrary{matrix}

\newcommand{\of}[1]{\left(#1\right)}
\newcommand{\off}[1]{\left[#1\right]}
\newcommand{\offf}[1]{\left\{#1\right\}}
\newcommand{\abs}[1]{\left|#1\right|}

\begin{document}
		\pgfplotsset{every axis/.append style={
			line width=1pt,
			legend style={font=\large, at={(0.97,0.85)}}},
	} %
	\pgfplotsset{compat=1.13}
\begin{acronym}

\acro{5G-NR}{5G New Radio}
\acro{3GPP}{3rd Generation Partnership Project}
\acro{AC}{address coding}
\acro{ACF}{autocorrelation function}
\acro{ACR}{autocorrelation receiver}
\acro{ADC}{analog-to-digital converter}
\acrodef{aic}[AIC]{Analog-to-Information Converter}     
\acro{AIC}[AIC]{Akaike information criterion}
\acro{aric}[ARIC]{asymmetric restricted isometry constant}
\acro{arip}[ARIP]{asymmetric restricted isometry property}

\acro{ARQ}{automatic repeat request}
\acro{AUB}{asymptotic union bound}
\acrodef{awgn}[AWGN]{Additive White Gaussian Noise}     
\acro{AWGN}{additive white Gaussian noise}

\acro{APSK}[PSK]{asymmetric PSK} 

\acro{waric}[AWRICs]{asymmetric weak restricted isometry constants}
\acro{warip}[AWRIP]{asymmetric weak restricted isometry property}
\acro{BCH}{Bose, Chaudhuri, and Hocquenghem}        
\acro{BCHC}[BCHSC]{BCH based source coding}
\acro{BEP}{bit error probability}
\acro{BFC}{block fading channel}
\acro{BG}[BG]{Bernoulli-Gaussian}
\acro{BGG}{Bernoulli-Generalized Gaussian}
\acro{BPAM}{binary pulse amplitude modulation}
\acro{BPDN}{Basis Pursuit Denoising}
\acro{BPPM}{binary pulse position modulation}
\acro{BPSK}{binary phase shift keying}
\acro{BPZF}{bandpass zonal filter}
\acro{BU}[BU]{Bernoulli-uniform}
\acro{BER}{bit error rate}
\acro{BS}{base station}
\acro{BC}{backscatter communication}

\acro{CP}{Cyclic Prefix}
\acrodef{cdf}[CDF]{cumulative distribution function}   
\acro{CDF}{cumulative distribution function}
\acrodef{c.d.f.}[CDF]{cumulative distribution function}
\acro{CCDF}{complementary cumulative distribution function}
\acrodef{ccdf}[CCDF]{complementary CDF}               
\acrodef{c.c.d.f.}[CCDF]{complementary cumulative distribution function}
\acro{CD}{cooperative diversity}

\acro{CDMA}{Code Division Multiple Access}
\acro{ch.f.}{characteristic function}
\acro{CIR}{channel impulse response}
\acro{cosamp}[CoSaMP]{compressive sampling matching pursuit}
\acro{CR}{cognitive radio}
\acro{cs}[CS]{compressed sensing}                   
\acrodef{cscapital}[CS]{Compressed sensing} 
\acrodef{CS}[CS]{compressed sensing}
\acro{CSS}{chirp spread spectrum}
\acro{CSI}{channel state information}
\acro{CCSDS}{consultative committee for space data systems}
\acro{CC}{convolutional coding}
\acro{Covid19}[COVID-19]{Coronavirus disease}
\acro{SSBC}{spectrum sharing backscatter communications}
\acro{CW}{continuous wave}
\acro{COT}{commercial off-the-shelf}

\acro{DAA}{detect and avoid}
\acro{DAB}{digital audio broadcasting}
\acro{DCT}{discrete cosine transform}
\acro{DDS}{direct digital synthesis}
\acro{dft}[DFT]{discrete Fourier transform}
\acro{DR}{distortion-rate}
\acro{DS}{direct sequence}
\acro{DS-SS}{direct-sequence spread-spectrum}
\acro{DTR}{differential transmitted-reference}
\acro{DVB-H}{digital video broadcasting\,--\,handheld}
\acro{DVB-T}{digital video broadcasting\,--\,terrestrial}
\acro{DL}{downlink}
\acro{DSSS}{Direct Sequence Spread Spectrum}
\acro{DFT-s-OFDM}{Discrete Fourier Transform-spread-Orthogonal Frequency Division Multiplexing}
\acro{DAS}{distributed antenna system}
\acro{DNA}{Deoxyribonucleic Acid}

\acro{EC}{European Commission}
\acro{EED}[EED]{exact eigenvalues distribution}
\acro{EIRP}{Equivalent Isotropically Radiated Power}
\acro{ELP}{equivalent low-pass}
\acro{eMBB}{enhanced mobile broadband}
\acro{EMF}{electric and magnetic fields}
\acro{EU}{European union}
\acro{EVT}{extreme value theorem}
\acro{ETSI}{European Telecommunications Standards Institute}

\acro{FC}[FC]{fusion center}
\acro{FCC}{Federal Communications Commission}
\acro{FD}{full-duplex}
\acro{FEC}{forward error correction}
\acro{FER}{frame error rate}
\acro{FFT}{fast Fourier transform}
\acro{FH}{frequency-hopping}
\acro{FH-SS}{frequency-hopping spread-spectrum}
\acrodef{FS}{Frame synchronization}
\acro{FSsmall}[FS]{frame synchronization}  
\acro{FDMA}{Frequency Division Multiple Access}

\acro{GA}{Gaussian approximation}
\acro{GF}{Galois field }
\acro{GG}{Generalized-Gaussian}
\acro{GIC}[GIC]{generalized information criterion}
\acro{GLRT}{generalized likelihood ratio test}
\acro{GPS}{Global Positioning System}
\acro{GMSK}{Gaussian minimum shift keying}
\acro{GSMA}{Global System for Mobile communications Association}

\acro{HAP}{high altitude platform}
\acro{HD}{half-duplex}

\acro{IDR}{information distortion-rate}
\acro{IFFT}{inverse fast Fourier transform}
\acro{iht}[IHT]{iterative hard thresholding}
\acro{i.i.d.}{independent, identically distributed}
\acro{IoT}{Internet of Things}                      
\acro{IR}{impulse radio}
\acro{lric}[LRIC]{lower restricted isometry constant}
\acro{lrict}[LRICt]{lower restricted isometry constant threshold}
\acro{ISI}{intersymbol interference}
\acro{ITU}{International Telecommunication Union}
\acro{ICNIRP}{International Commission on Non-Ionizing Radiation Protection}
\acro{IEEE}{Institute of Electrical and Electronics Engineers}
\acro{ICES}{IEEE international committee on electromagnetic safety}
\acro{IEC}{International Electrotechnical Commission}
\acro{IARC}{International Agency on Research on Cancer}
\acro{IS-95}{Interim Standard 95}

\acro{LB}{LoRa backscatter}
\acro{LEO}{low earth orbit}
\acro{LF}{likelihood function}
\acro{LLF}{log-likelihood function}
\acro{LLR}{log-likelihood ratio}
\acro{LLRT}{log-likelihood ratio test}
\acro{LOS}{Line-of-Sight}
\acro{LRT}{likelihood ratio test}
\acro{wlric}[LWRIC]{lower weak restricted isometry constant}
\acro{wlrict}[LWRICt]{LWRIC threshold}
\acro{LPWAN}{low power wide area networks}
\acro{LoRaWAN}{low power long range wide area network}
\acro{NLOS}{non-line-of-sight}

\acro{MB}{multiband}
\acro{MC}{multicarrier}
\acro{MDS}{mixed distributed source}
\acro{MF}{matched filter}
\acro{m.g.f.}{moment generating function}
\acro{MI}{mutual information}
\acro{MIMO}{multiple-input multiple-output}
\acro{MISO}{multiple-input single-output}
\acrodef{maxs}[MJSO]{maximum joint support cardinality}                       
\acro{ML}[ML]{maximum likelihood}
\acro{MMSE}{minimum mean-square error}
\acro{MMV}{multiple measurement vectors}
\acrodef{MOS}{model order selection}
\acro{M-PSK}[${M}$-PSK]{$M$-ary phase shift keying}                       
\acro{M-APSK}[${M}$-PSK]{$M$-ary asymmetric PSK} 
\acro{MTC}{machine type communication}
\acro{MGF}{moment generating function} 
\acro{M-QAM}[$M$-QAM]{$M$-ary quadrature amplitude modulation}
\acro{MRC}{maximal ratio combiner}                  
\acro{maxs}[MSO]{maximum sparsity order}                                      
\acro{M2M}{machine to machine}                                                
\acro{MUI}{multi-user interference}
\acro{mMTC}{massive machine type communications}      
\acro{mm-Wave}{millimeter-wave}
\acro{MP}{mobile phone}
\acro{MPE}{maximum permissible exposure}
\acro{MAC}{media access control}
\acro{NB}{narrowband}
\acro{NBI}{narrowband interference}
\acro{NLA}{nonlinear sparse approximation}
\acro{NLOS}{Non-Line of Sight}
\acro{NTIA}{National Telecommunications and Information Administration}
\acro{NTP}{National Toxicology Program}
\acro{NHS}{National Health Service}
\acro{NB-IoT}{narrowband Internet of things}

\acro{OC}{optimum combining}                             
\acro{OC}{optimum combining}
\acro{ODE}{operational distortion-energy}
\acro{ODR}{operational distortion-rate}
\acro{OFDM}{orthogonal frequency-division multiplexing}
\acro{OOK}{ON-OFF Keying}
\acro{omp}[OMP]{orthogonal matching pursuit}
\acro{OSMP}[OSMP]{orthogonal subspace matching pursuit}
\acro{OQAM}{offset quadrature amplitude modulation}
\acro{OQPSK}{offset QPSK}
\acro{OFDMA}{Orthogonal Frequency-division Multiple Access}
\acro{OPEX}{Operating Expenditures}
\acro{OQPSK/PM}{OQPSK with phase modulation}

\acro{PAM}{pulse amplitude modulation}
\acro{PAR}{peak-to-average ratio}
\acrodef{pdf}[PDF]{probability density function}                      
\acro{PDF}{probability density function}
\acrodef{p.d.f.}[PDF]{probability distribution function}
\acro{PDP}{power dispersion profile}
\acro{PMF}{probability mass function}                             
\acrodef{p.m.f.}[PMF]{probability mass function}
\acro{PN}{pseudo-noise}
\acro{PPM}{pulse position modulation}
\acro{PRake}{Partial Rake}
\acro{PSD}{power spectral density}
\acro{PSEP}{pairwise synchronization error probability}
\acro{PSK}{phase shift keying}
\acro{PD}{power density}
\acro{8-PSK}[$8$-PSK]{$8$-phase shift keying}
\acro{PR}{primary receiver}
\acro{PT}{primary trasmitter}
 
\acro{FSK}{frequency shift keying}

\acro{QAM}{Quadrature Amplitude Modulation}
\acro{QPSK}{quadrature phase shift keying}
\acro{OQPSK/PM}{OQPSK with phase modulator }

\acro{RD}[RD]{raw data}
\acro{RDL}{"random data limit"}
\acro{RFEH}{RF energy harvesting}
\acro{ric}[RIC]{restricted isometry constant}
\acro{rict}[RICt]{restricted isometry constant threshold}
\acro{rip}[RIP]{restricted isometry property}
\acro{ROC}{receiver operating characteristic}
\acro{rq}[RQ]{Raleigh quotient}
\acro{RS}[RS]{Reed-Solomon}
\acro{RSC}[RSSC]{RS based source coding}
\acro{r.v.}{random variable}              
\acro{R.V.}{random vector}
\acro{RMS}{root mean square}
\acro{RFR}{radiofrequency radiation}
\acro{RIS}{reconfigurable intelligent surface}
\acro{RNA}{RiboNucleic Acid}
\acro{Rx}{receiver}

\acro{SA}[SA-Music]{subspace-augmented MUSIC with OSMP}
\acro{SCBSES}[SCBSES]{Source Compression Based Syndrome Encoding Scheme}
\acro{SCM}{sample covariance matrix}
\acro{SEP}{symbol error probability}
\acro{SER}{symbol error rate}
\acro{SG}[SG]{sparse-land Gaussian model}
\acro{SIMO}{single-input multiple-output}
\acro{SINR}{signal-to-interference plus noise ratio}
\acro{SIR}{signal-to-interference ratio}
\acro{SISO}{single-input single-output}
\acro{SMV}{single measurement vector}
\acro{SNR}[\textrm{SNR}]{signal-to-noise ratio} 
\acro{sp}[SP]{subspace pursuit}
\acro{SS}{spread spectrum}
\acro{SW}{sync word}
\acro{SAR}{specific absorption rate}
\acro{SSB}{synchronization signal block}
\acro{SR}{secondary receiver}
\acro{ST}{secondary trasmitter}
\acro{SF}{spreading factor}

\acro{TH}{time-hopping}
\acro{ToA}{time-of-arrival}
\acro{TR}{transmitted-reference} 
\acro{TW}{Tracy-Widom}
\acro{TWDT}{TW Distribution Tail}
\acro{TCM}{trellis coded modulation}
\acro{TDD}{time-division duplexing}
\acro{TDMA}{time division multiple access}
\acro{Tx}{transmitter}

\acro{UAV}{unmanned aerial vehicle}
\acro{uric}[URIC]{upper restricted isometry constant}
\acro{urict}[URICt]{upper restricted isometry constant threshold}
\acro{UWB}{ultrawide band}
\acro{UWBcap}[UWB]{Ultrawide band}   
\acro{URLLC}{Ultra Reliable Low Latency Communications}
         
\acro{VCO}{voltage-controlled oscillator}    
\acro{wuric}[UWRIC]{upper weak restricted isometry constant}
\acro{wurict}[UWRICt]{UWRIC threshold}                
\acro{UE}{user equipment}
\acro{UL}{uplink}
\acro{URLLC}{ultra reliable low latency communications}

\acro{WiM}[WiM]{weigh-in-motion}
\acro{WLAN}{wireless local area network}
\acro{wm}[WM]{Wishart matrix}                               
\acroplural{wm}[WM]{Wishart matrices}
\acro{WMAN}{wireless metropolitan area network}
\acro{WPAN}{wireless personal area network}
\acro{wric}[WRIC]{weak restricted isometry constant}
\acro{wrict}[WRICt]{weak restricted isometry constant thresholds}
\acro{wrip}[WRIP]{weak restricted isometry property}
\acro{WSN}{wireless sensor network}                        
\acro{WSS}{wide-sense stationary}
\acro{WHO}{World Health Organization}
\acro{Wi-Fi}{wireless fidelity}

\acro{sss}[SpaSoSEnc]{sparse source syndrome encoding}

\acro{VLC}{visible light communication}
\acro{VPN}{virtual private network} 
\acro{RF}{radio frequency}
\acro{FSO}{free space optics}
\acro{IoST}{Internet of space things}

\acro{GSM}{Global System for Mobile Communications}
\acro{2G}{second-generation cellular networks}
\acro{3G}{third-generation cellular networks}
\acro{4G}{fourth-generation cellular networks}
\acro{5G}{5th-generation cellular networks}	
\acro{gNB}{next generation node B base station}
\acro{NR}{New Radio}
\acro{UMTS}{Universal Mobile Telecommunications Service}
\acro{LTE}{Long Term Evolution}

\acro{QoS}{Quality of Service}
\end{acronym}

\title{LoRa Backscatter Communications: Temporal, Spectral, and Error Performance Analysis}


\author{Ganghui Lin,~\IEEEmembership{Graduate Student Member,~IEEE} Ahmed Elzanaty,~\IEEEmembership{Senior Member,~IEEE}, and \\Mohamed-Slim Alouini,~\IEEEmembership{Fellow,~IEEE}
\thanks{Ganghui Lin and Mohamed-Slim Alouini are with the Division of Computer, Electrical and Mathematical Sciences and Engineering, King Abdullah University of Science and Technology, Thuwal 23955-6900, Saudi Arabia (e-mail: ganghui.lin@kaust.edu.sa, slim.alouini@kaust.edu.sa).}
\thanks{A. Elzanaty is with the 5GIC \& 6GIC, Institute for Communication Systems (ICS), University of Surrey, Guildford, GU2 7XH, United Kingdom (e-mail: a.elzanaty@surrey.ac.uk).}
\thanks{The source codes can be accessed at https://github.com/SlinGovie/LoRa-Backscatter-Performance-Analysis.
}
}
\maketitle
\begin{abstract}
 \acf{LB} communication systems can be considered as a potential candidate for ultra \ac{LPWAN} because of their low cost and low power consumption. In this paper, we comprehensively analyze \ac{LB} modulation from various aspects, i.e., temporal, spectral, and error performance characteristics. First, we propose a signal model for  \ac{LB} signals that accounts for the limited number of loads in the tag. 
Then, we investigate the spectral properties of \ac{LB}  signals, obtaining a closed-form expression for the power spectrum.  Finally, we derived the \ac{SER} of \ac{LB} with two decoders, i.e., the \ac{ML} and \ac{FFT} decoders, in both \ac{AWGN} and double Nakagami-m fading channels. The spectral analysis shows that out-of-band emissions for \ac{LB} satisfy the \ac{ETSI} regulation only when considering a relatively large number of loads. {For the error performance, unlike conventional {LoRa}, the \ac{FFT} decoder is not optimal. Nevertheless, the \ac{ML} decoder can achieve a performance similar to conventional {LoRa} with a moderate number of loads.} 
\end{abstract}
\begin{IEEEkeywords} 
		\acf{SER}; \acf{LB}; \acf{IoT}; power spectral density
\end{IEEEkeywords}
	\acresetall 

\section{Introduction}
The goal of \acs{5G} and beyond networks is to realize three core services, i.e., \ac{eMBB}, \ac{URLLC}, and \ac{mMTC} \cite{BocHeathPopovski:14,ShuppingAminSlim:20},  to support extreme network capacity, latency-sensitive  critical missions, and a massive number of connected devices, respectively. 
In \ac{mMTC}, a colossal variety of inexpensive devices is needed to communicate at low power consumption. With nearly 30 billion devices expected to be connected by 2030 \cite{vailshery_2022}, the prospect of \ac{mMTC} makes it possible for \ac{IoT} applications such as smart cities, homes, and agriculture.

In this regard, many wireless technologies have recently emerged to realize such low-power communications with low-priced devices. For instance, \ac{LoRaWAN}, SigFoX, and \ac{NB-IoT} are three prominent technologies supporting \ac{LPWAN}  \cite{AEloramodulation,RazKulSoo:17,PiaElzGioChi:17b,s23031698}. 
However, some application scenarios put forward more stringent requirements on the power consumption that even technologies such as \ac{LPWAN} cannot meet.

In this context, \ac{BC} is one of the most remarkable technologies for ultra-low-power communications with its ability to operate with sub-milliwatt power consumption. Typical \ac{BC} systems contain backscatter tags and transceivers. The tag reflects the incoming excitation signals while modulating it with its data by changing the incident signal amplitude, frequency, or phase \cite{RezTelHer:20}. The innovation of \ac{BC} lies in its radio-passive property that removes the power-consuming analog components such as RF oscillators, decoupling capacitors, and crystals. Hence, \ac{BC} devices can operate with emerging flexible and printed batteries or even harvest environmental energy\cite{advancesBSc2021,VanHoaKim:18}. 

\ac{BC} can be divided into three categories based on the adopted architecture:  \textit{(i)  monostatic backscatter}, where a  tag modulates an excitation signal generated by a transceiver, then reflects it back to the transceiver; \textit{(ii) bistatic backscatter},  where the \ac{Tx} and \ac{Rx} are separated;  \textit{(iii)  ambient backscatter}, where the excitation signal comes from existing surrounding RF sources in the environment \cite{WanGaiTel:16,LiuSmith:13}.

Despite the low-cost, low-power merits that \ac{BC} owns, one of the most detrimental defects of \ac{BC} is its limited communication range. The electromagnetic waves suffer from severe multipath fading and strong attenuation in urban or indoor environments, significantly reducing the coverage of \ac{BC} \cite{9455142}. In some medical or wearable applications, the human body also shortens the communication range\cite{4pbsk}. 

One possible solution to this problem is \ac{LB}, where the backscatter tag generates \ac{CSS} modulated signals to improve the robustness of \ac{BC}\cite{3lorabsk}. LoRa modulated signals have exceptionally high sensitivity up to $-149$ dBm, compensating for the long-range path loss as well as the multipath fading effects. Although \ac{LB} is similar, to some extent, to LoRa modulation in terms of high sensitivity, they differ in many aspects. First, the \ac{LB} modulated signals only have a finite number of phases because of the limited number of antenna loads in the backscatter tags.  Second, conventional LoRa \acp{Tx} are active devices that directly transmit  signals  to the gateway, hopefully through a line of sight link. In contrast, in \ac{LB} the signals from the \ac{Tx}  pass through the channel between the \ac{Tx} and tag before backscattered to the \ac{Rx}, i.e., double-fading channel. 

The aforementioned aspects make a huge difference in analyzing the performance of \ac{LB}. More precisely, a double fading channel should be considered in \ac{LB}, resulting in a more complicated \ac{PDF} for the channel amplitude. The discrete phases also make the waveforms representing different symbols non-orthogonal at the Nyquist sampling rate, rendering non-equivalency between the \ac{ML} and \ac{FFT} decoders.


\subsection{Related work}
In the following, we review the state-of-the-art \ac{LB} schemes and the performance analysis regarding \ac{BC} and LoRa as well as \ac{LB}.

\subsubsection{The State-of-the-art of \ac{LB}}
In order to increase the range for \ac{BC}, a \ac{LB} communication system with a harmonic cancellation scheme is proposed in \cite{3lorabsk}. The proposed system consists of a single-tone carrier \ac{Tx}, a backscatter tag, and a \ac{Rx}. The tag uses the single tone to synthesize \ac{CSS} modulated signals and reflect them to the \ac{Rx}. To avoid interference from the RF source, a frequency shift is introduced in the tag by multiplying the incoming signals by an approximated cosine and sine with discontinuous step transitions, resulting in high-frequency components, i.e., the harmonics. The harmonic cancellation is achieved by adding more voltage levels, i.e., more antenna loads in the tag, to better approximate a pure sinusoid. 
The design presents a reliable wide area network coverage, i.e., $475$~m from the \ac{Tx} and \ac{Rx},  provided by a low-cost device that consumes $1000$x lower power than a normal LoRa \ac{Tx}. However, the tag still needs a battery as the power source rather than harvesting energy from the environment. 
In \cite{selfsustainlora2021}, an \ac{RFEH} \ac{LB} scheme is proposed. It is reported that the tag can self-start up while harvesting RF energy from the excitation signal as low as $-22.5$ dBm and send the acquired data to a \ac{Rx} that is $381$ m away. There are two possible reasons for the shorter range compared to \cite{3lorabsk}. 
One is that the \ac{RFEH} LoRa tag only uses two antenna loads, resulting in more harmonics. Another reason is that the \ac{RFEH} LoRa tag harvest energy from the RF source instead of having an embedded battery so that a proportion of the energy is used to power the IC. 
To resolve the problem of deployment difficulties of traditional \acl{HD} \ac{LB}, the first \acl{FD} \ac{LB} architecture is proposed in \cite{fullduplexlorabsc}. Nevertheless, the \acl{FD} \ac{LB} achieves a shorter range due to a different sensitivity protocol and a reduced link budget introduced by a hybrid coupler architecture used to reduce cost.

The above-mentioned \ac{LB} systems all use single-tone carriers and the tag modulates its bits into a subcarrier to generate LoRa packets. However, there are other \ac{LB} systems that use normal LoRa modulated waves as excitation signals \cite{xorlora,plora,freeback2021,onoffkeylorabsc,wearableLoRa2020}.  These ambient backscatter designs use simpler modulation methods such as \ac{OOK}, which leads to less complicated tag design at a cost of either shorter communication range, more complicated decoding algorithms, or lower throughput. 

\subsubsection{Performance Analysis for \ac{BC}, LoRa, and \ac{LB}}
{For \ac{BC}, several works have considered its performance analysis in terms of detection schemes, error rate, and information rate. In \cite{7059230}, the authors consider \ac{FSK} with a coherent receiver for the bistatic backscatter radio channel. Nevertheless,  coherent detection is complex and requires \ac{CSI} estimation. To avoid this difficulty, non-coherent detection schemes for ambient \ac{BC} have been considered in \cite{7769255,8532293,8721108}. 
With the different detection schemes mentioned above, the exact \ac{BER} for the ambient backscatter system is derived in \cite{8532293}.  Since it uses unknown ambient signals as carriers, the actual capacity is of interest. In \cite{8721108}, the achievable rate of the ambient \ac{BC} system is analyzed under different channels. Although the performance analysis of several \ac{BC} systems with conventional modulation schemes (e.g., \ac{BPSK}, \ac{FSK}, \ac{OOK}) have been analyzed in the literature, few works consider long-range \ac{CSS} modulation such as \ac{LB}. In the following, we first discuss the performance analysis of conventional LoRa then \ac{LB}. }


For conventional LoRa,
the power spectrum characteristics of normal LoRa are analyzed in terms of the Fresnel functions\cite{AEloramodulation}, in which the derived expressions are not applicable for \ac{LB} because of its limited number of phases.
In \cite{loraphysical2018}, the \ac{BER} of LoRa in an \ac{AWGN} channel is derived by using the Monte-Carlo approximation, but the method requires lots of computing resources.
The exact \ac{SER} of orthogonal signaling with non-coherent detection is presented in \cite[eq: 4.5-44]{proakis2007digital}, which is applicable for LoRa. However, the equation is only derived in an \ac{AWGN} scenario and the high-order combination involved in the equation will introduce severe numerical problems.
The exact \ac{SER} of LoRa modulation in various fading scenarios, i.e., Rayleigh, Rician, and Nakagami fading channels, are analyzed in \cite{berLoRaExactSensor}. Nevertheless, the expressions are hardly computable because of the remaining high-order combination.
In \cite{asymptoticBERLoRa2021}, simple asymptotic \ac{BER} expressions of LoRa modulation over Rician and Rayleigh channels are derived at cost of accuracy. The numerical results show poor accuracy at low \ac{SNR}.
In \cite{7closedform}, an accurate closed-form approximation of the \ac{BER} in both \ac{AWGN} and Rayleigh fading channels is derived. However, the analysis for other fading models such as Rician and Nakagami-m fading channels is not included.
In \cite{computablelora2021}, the author proposed a new approach based on the Marcum function to estimate the \ac{BER} of LoRa in different propagation environments, namely, \ac{AWGN}, Rayleigh, Rician, and Nakagami channels. 
{In \cite{baruffa2020error}, a  modification of a union bound on the error probability is proposed to calculate the \ac{BER} with less complexity. The approximation is applicable for both coherent and non-coherent detection. The theoretical performance analysis is also extended to Hamming-coded LoRa systems.}
{In \cite{9693529}, a novel modulation scheme is introduced that employs both up-chirp and down-chirp simultaneously, and its \ac{BER} is derived in the \ac{AWGN} channel. Numerical results demonstrate that this scheme achieves a doubled throughput without a significant increase in BER.
Further scenarios have been studied such as channel coding\cite{loraTheoreticalAndExperimental}, interference channels \cite{lorainterference2019,loraSameSFInterference2018}, 
and imperfect orthogonality\cite{loraImperfectOrtho2018,Benkhalifa:22} for LoRa. Although the performance analysis of conventional LoRa modulation over \ac{AWGN} and various fading channels have been extensively studied in the literature,  they do not consider the double fading channel and the limited number of loads that characterize \ac{LB}.
}

In \cite{9641890}, the error performance of \ac{LB} is discussed in \ac{AWGN} channels. However, the essential characteristics of \ac{LB} for the signal model are not considered. For instance, the fact that  \ac{LB} modulated signals only have limited available phases is not considered. Also, since the performance was only in an \ac{AWGN} channel, it does not account for the double-fading effects which are crucial for \ac{LB}. 
\subsection{Contributions}
Although some prototypes have proved the feasibility of \ac{LB}, more theoretical insights are needed in order to support the design and deployment of \ac{LB} systems. Nevertheless,  a comprehensive performance analysis of \ac{LB} communication systems is still lacking in the literature. In fact, none of the aforementioned work has considered the main features that characterize \ac{LB} such as finite phases and the double fading channel.  

In this paper, we comprehensively analyze \ac{LB} modulation from various aspects, i.e., temporal, spectral, and error performance characteristics. More precisely, we provide the first expression for the \ac{LB} modulated signal that accounts for the reduced complexity of the system compared to LoRa. The reason is that the number of antenna loads is finite, limiting the number of available phases for the signal. 

Based on the provided signal expression, we analytically derive a closed-form expression for the power spectrum of baseband \ac{LB} signals. The derivation of the power spectral density involves integration over the non-linear phase quantization function. The spectral analysis provides a better understanding of the adjacent channel interference for \ac{LB} modulation and its relation to the number of loads (i.e., system complexity) \cite{ETSI}.  

 Also, we derive the optimal decoder for \ac{LB}, i.e., \ac{ML}, and compare it with the conventional \ac{FFT} decoder. For \ac{ML} and \ac{FFT} decoder, we conduct error performance analysis in \ac{AWGN} and fading channels. More precisely, we derive closed-form approximations for the \ac{SER} in both \ac{AWGN} and double Nakagami-m fading channels for fixed transmit power. Besides, given a constraint on the average transmit power, we provide the optimal power allocation scheme using water-filling and analyze the corresponding average \ac{SER}.
 
 The performance analysis of \ac{LB} is quite involving compared to normal LoRa. For instance, unlike normal LoRa, the \ac{LB} waveforms representing different symbols are not orthogonal when sampled at the Nyquist rate. This leads to a \ac{SER} expression involving  a product of  Marcum Q functions with different shape parameters. Additionally, the \ac{PDF} of the cascaded Nakagami-m fading channel  entails integration over a modified Bessel function of the second kind, putting further the difficulty in numerical evaluation. 

The contribution of this paper can be summarized as follows:  
\begin{itemize}
\item We  provide a novel signal expression in the time-domain for \ac{LB} signal with a generalized number of loads.
\item  We derive a closed-form expression for the power spectrum of \ac{LB} signals and investigate whether \ac{LB} meets the regulation of \ac{ETSI} for adjacent channel interference. 
\item We provide a \ac{ML}-based decoder for \ac{LB}.
\item We derive closed-form approximations for the \ac{SER} of \ac{LB} in both \ac{AWGN} and double Nakagami-m fading channels, considering both \ac{FFT} and \ac{ML} decoders.
\item We study the \ac{SER} performance of \ac{LB} for various power allocation schemes, i.e., fixed and optimized power allocation techniques.
\end{itemize}
\subsection{Notations}
\begin{table}[t]
\centering
\caption{Table of Symbols}
\begin{tabular}{ccc}
	\toprule  
	Symbol & Meaning\\
	\midrule  
	$D$&decoder type\\
	$\sigma^2$& noise power per dimension\\
	$\gamma$ &  \ac{SNR}\\
	$E_\textrm{s}$, $\overline{E_\textrm{s}}$ & instantaneous, average symbol energy\\
	$a$, $i$& transmit symbol index, decoder output bin index\\
	$\mathcal{L}^{D}_{\left(a,i\right)}$ & $i$-th decoder output bin value for symbol $a$ \\
	$\hat{\mathcal{L}}^{D}_{\left(a\right)}$ & $\mathop{\max}\limits_{ i,i\neq a}\offf{\mathcal{L}^{D}_{\left(a,i\right)}} $\\
	$ \xi^{D}_{\left(a,i\right)} $& decoder outputs without noise\\
	$\mathcal{W}^D_{\of{a,i}}$&noise at the decoder\\
	$\mu_a$, $\sigma_a^2$ & expectation and variance of $\mathcal{L}^{D}_{\left(a,a\right)}$\\
	$d_1$, $d_2$ & link distances of the \ac{Tx}-tag and tag-\ac{Tx} links\\
	$m$, $\Omega$& shape and spread parameter of Nakagami distribution \\
	$r$, $v$, $n$& $m/\Omega$, $m_1+m_2$, and $|m_1-m_2|$ \\
	$x_t$, $\omega_t$, $N_\textrm{GH}$& points, weights, and number of points in GH quadrature\\
	$x_e$, $\omega_e$, $N_\textrm{GL}$& points, weights, and number of points in GL quadrature\\
	\bottomrule 
\end{tabular}
\label{table:parameters}
\end{table}
Throughout this paper, we denote the \ac{Tx}-tag  and tag-\ac{Rx} links with subscripts 1 and 2, respectively,  
vectors with bold small letters, 
\acp{r.v.} with calligraphic letters, e.g., $\mathcal{L}$, 
the \ac{PDF} of $\mathcal{L}$ with $f_{\mathcal{L}}\of{l}$, 
the \ac{CDF} of $\mathcal{L}$ with $F_{\mathcal{L}}\of{l}$,  
the expectation and variance of $\mathcal{L}$ with $\mathbb{E}\off{\mathcal{L}}$ and $\mathbb{V}\off{\mathcal{L}}$, respectively, complex
Gaussian distribution that has mean $\mu$ and variance $\sigma^2$ with $\mathcal{CN}\left(\mu,\sigma^2\right)$,  
complex conjugate of a complex variable $x$ with  $x^{*}$, 
and inner product of vectors $\boldsymbol{\alpha}$ and $\boldsymbol{\beta}$ with $<\boldsymbol{\alpha},\boldsymbol{\beta}>$. 
The main symbols considered in this paper are listed in Table~\ref{table:parameters}.
\subsection{Structure of the paper}
The remainder of this paper is structured as follows. First, the system model is presented in Section~\ref{sec:systemmodel}, and the spectral analysis of \ac{LB} is provided in Section~\ref{sec:SpectralAnalysis}. In Section~\ref{sec:performance}, the error performance in \ac{AWGN} and fading channels are investigated. Then, the numerical results are presented in Section~\ref{sec:numerical}, and Section~\ref{sec:conclude} concludes this paper.
\section{System Setup and Signal Model}
\label{sec:systemmodel}
In this section, we first explain the system setup and channel model for \ac{LB} communication systems. Then, we provide the analytical expressions for \ac{LB} signals, and we propose two decoders for \ac{LB}.
\subsection{System Setup and Channel Model}
\begin{figure}[t]
	\psfrag{X}[c][c][0.8]{$\boldsymbol{x_a}\off{k}$}
	\psfrag{Tx}[c][c][0.8]{Transmitter}
	\psfrag{Rx}[c][c][0.8]{Receiver}
	\psfrag{BT}[c][c][0.8]{Backscatter Tag}
	\psfrag{STE}[c][c][0.8]{Single-tone excitation}
	\psfrag{LP}[c][c][0.8]{LoRa packets}
	\psfrag{h1}[c][c][0.8]{$h_1$}
	\psfrag{h2}[c][c][0.8]{$h_2$}
	\centering
	\includegraphics[width=0.99\linewidth]{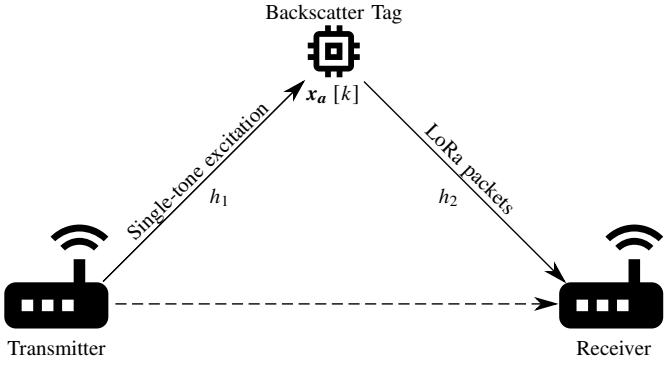}
	\caption{Considered \ac{LB} communication system.}
	\label{fig:systemdiagram}
\end{figure}
We consider a \ac{LB} communication system with a \ac{Tx} generating a carrier wave, a backscatter tag, and a \ac{Rx}, as shown in Fig.~\ref{fig:systemdiagram}. The \ac{Tx} sends a single-tone excitation wave to the tag with carrier frequency $f_{\textrm{c}}$. Then, the tag modulates the excitation signal with its data using \ac{CSS} modulation and reflects it to the \ac{Rx}. We consider a scheme where the tag introduces a frequency offset $\Delta f$ to the incoming signal, shifting it to a channel centered at $f_{\textrm{c}}+\Delta f$\footnote{{$\Delta f$ is typically $3$ orders of magnitude lower than $f_{\text{c}}$.}}to avoid interference from the direct link between \ac{Tx} and \ac{Rx} \cite{3lorabsk}. {\ac{LB} networks can be scaled by considering different SFs and/or carrier frequencies with negligible  interference among users.}

Also, we consider a flat double fading channel, where the complex channel gains of the \ac{Tx}-tag and tag-\ac{Rx} links are represented by $h_1$ and $h_2$, respectively. 
The received signal, sampled at the Nyquist rate, can be expressed as
\begin{equation}
	\label{eq:receivedsignal}
	{r_a}\left[k\right]=h_\textrm{1}\,h_\textrm{2}\,\sqrt{E_{\textrm{s}}}\,{x_a}\left[k\right]+{\omega}\left[k\right], \,\forall k \in \{0,1,\cdots,M-1\},
\end{equation}
where $E_{\textrm{s}}$ denotes the baseband symbol energy, $x_a[k]$ denotes $k^\textrm{th}$ sample of the baseband \ac{LB} signal with data symbol $a$, and ${\omega}\left[k\right]$ represents the thermal noise drawn from $\mathcal{CN}\left(0,2\sigma^2\right)$. 
For simplicity, the multiplication of two channel gains can be expressed in an overall channel gain $h$, i.e., $h \triangleq h_1h_2$. 
{This channel model includes the case of monostatic \ac{BC} by considering $h_1=h_2=\sqrt{h}$.}

In the following, we explain the backscatter process and provide analytical expressions for \ac{LB} baseband signals.

	\subsection{Backscatter Process and Signal Model}
	The reflected LoRa signals are synthesized by the RF switch that toggles between different antenna loads to create a set of reflection coefficients. 
	The set of coefficients is selected to change the amplitude and phase of the incoming signal $S_\textrm{in}$. More precisely, let us define  $\Gamma_\textrm{L}$ and $\Gamma_\textrm{A}$ as the load impedance and antenna impedance, respectively. The reflected signal $S_{\textrm{out}}$ can be expressed as 
	\begin{equation}
		S_{\textrm{out}}\left(t\right)=\frac{\Gamma_\textrm{L}\left(t\right)-\Gamma_\textrm{A}}{\Gamma_\textrm{L}\left(t\right)+\Gamma_\textrm{A}}\, S_{\textrm{in}}\left(t\right)=\left|\Gamma_\textrm{T}\left(t\right)\right|e^{j\theta_\textrm{T}\left(t\right)}\, S_{\textrm{in}}\left(t\right),
	\end{equation}
	where the change of $\Gamma_\textrm{L}\of{t}$ is discrete so that the possible number of phases $\theta_\textrm{T}\of{t}$ and amplitudes $\left|\Gamma_\textrm{T}\left(t\right)\right|$ are limited. However, to perfectly synthesize LoRa signals, we need an infinite number of phases so that the backscattered signals are only approximations of the ideal sinusoidal waves. For this reason, increasing the number of phases in \ac{LB} can enhance the system performance and approach traditional LoRa systems.
	On the other hand, increasing the number of possible phases raises not only the design complexity by adding more loads, but also the switching  frequency, reducing the tag service life. 
	
	\begin{figure}[t]
		\psfrag{Quadrature}[c][c][0.7]{Quadrature}
		\psfrag{In-phase}[c][c][0.7]{In-phase}
		\psfrag{Time}[c][c][0.7]{Time}
		\psfrag{Amplitude}[c][c][0.7]{Amplitude}
		\centering 
		\subfloat[]{
			\label{Fig:phase_dia}
			\includegraphics[width=0.31\linewidth]{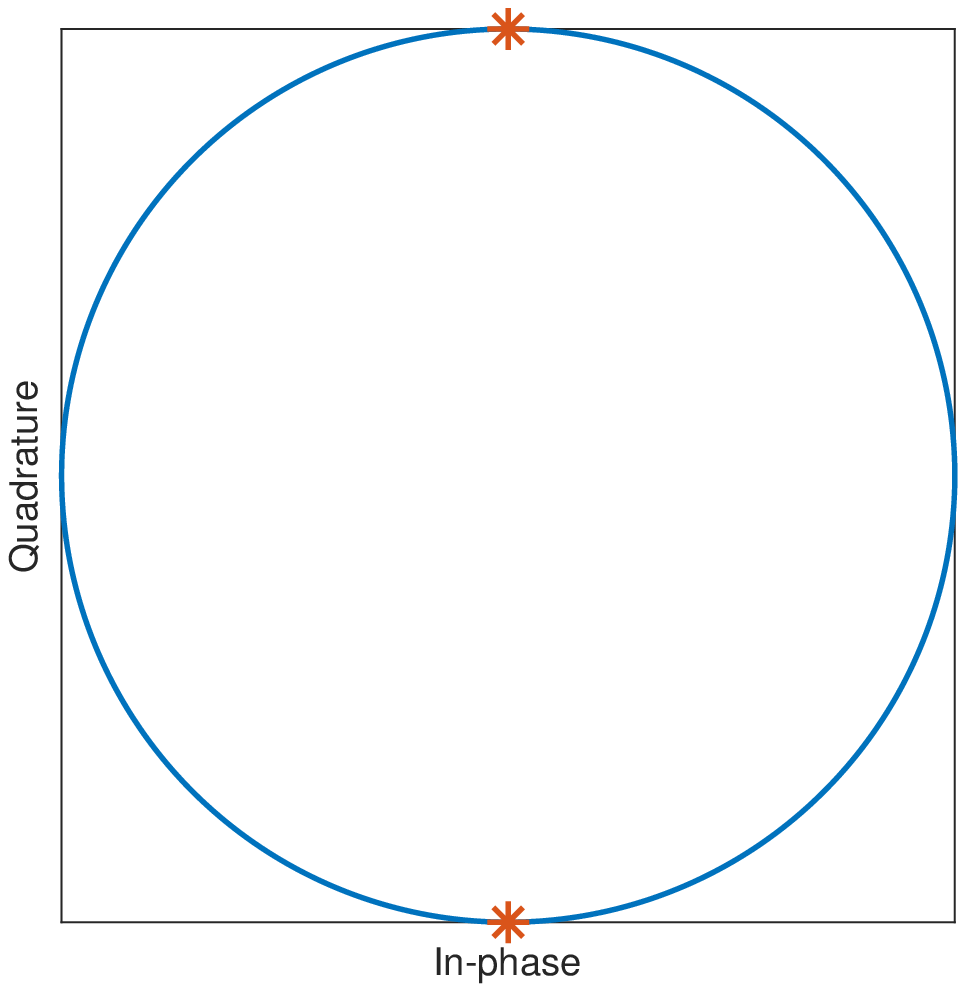}
			\includegraphics[width=0.31\linewidth]{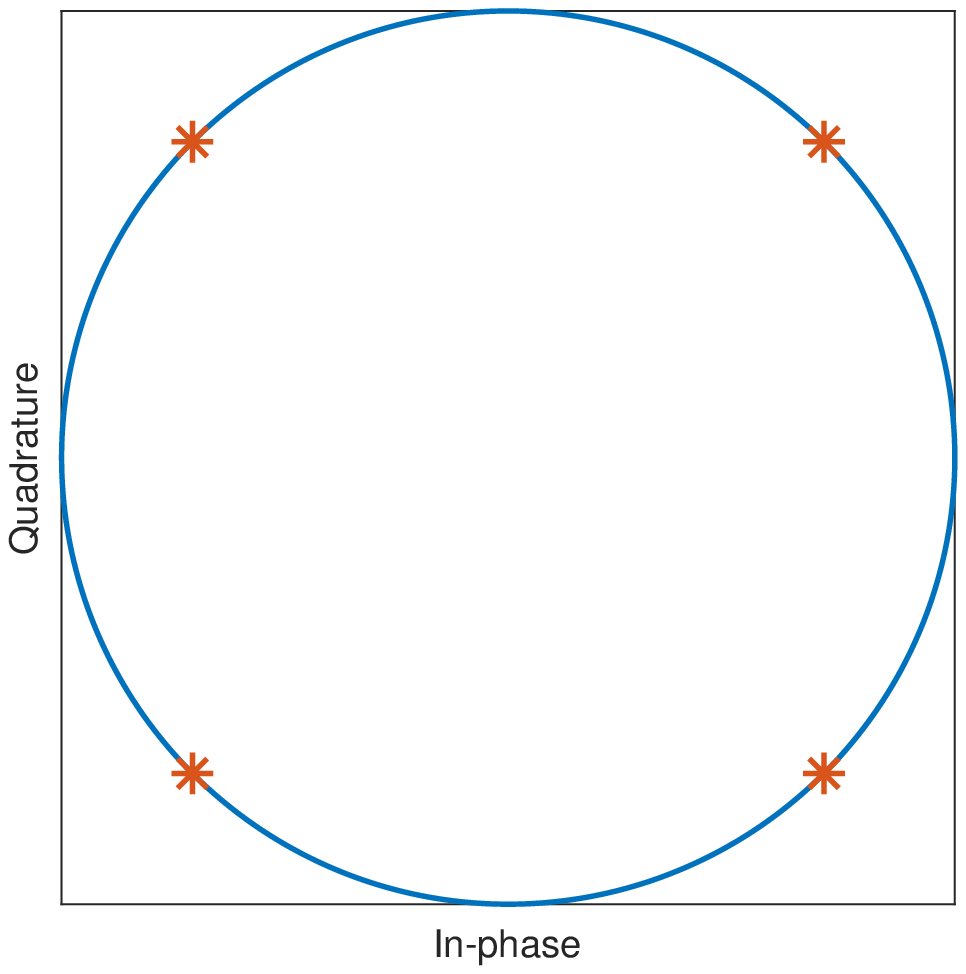}
			\includegraphics[width=0.31\linewidth]{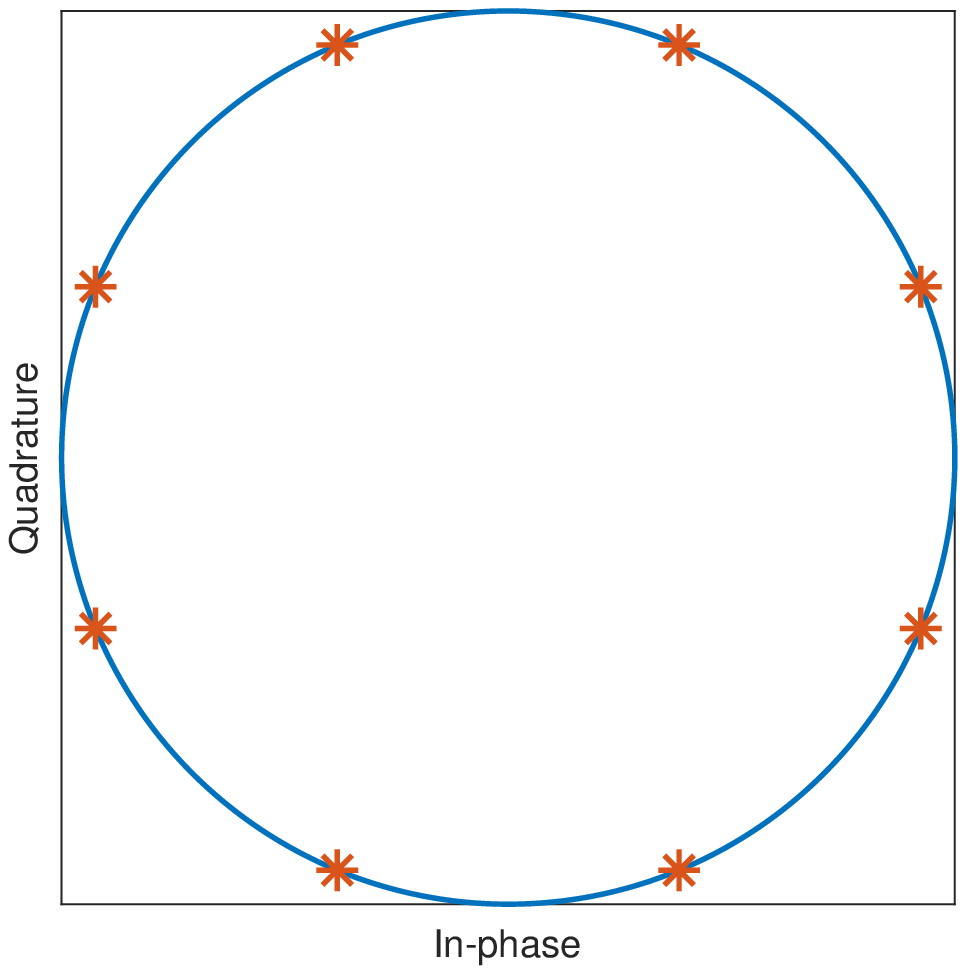}
		}
		\qquad
		\subfloat[]{
			\label{Fig:inphase}
			\includegraphics[width=0.31\linewidth]{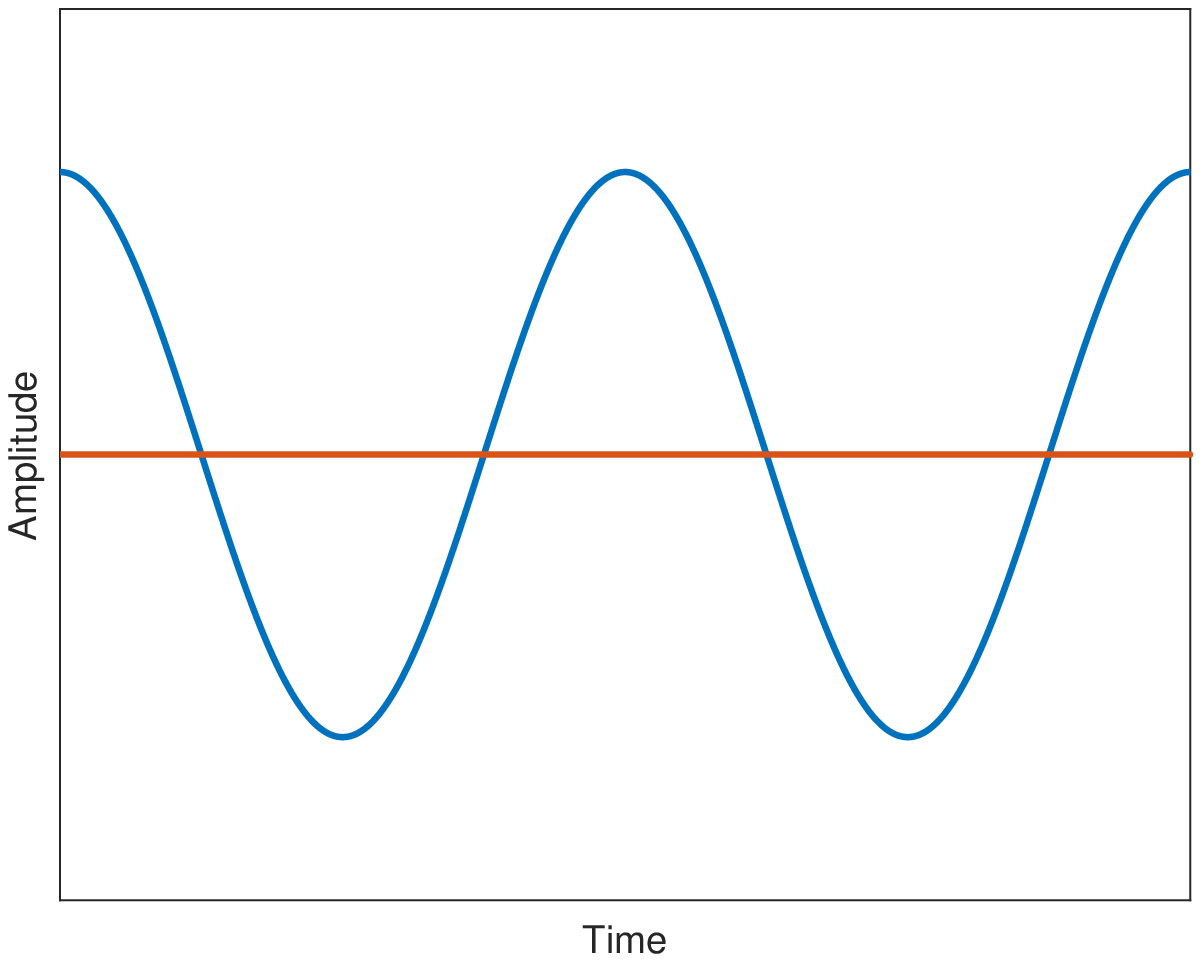}
			\includegraphics[width=0.31\linewidth]{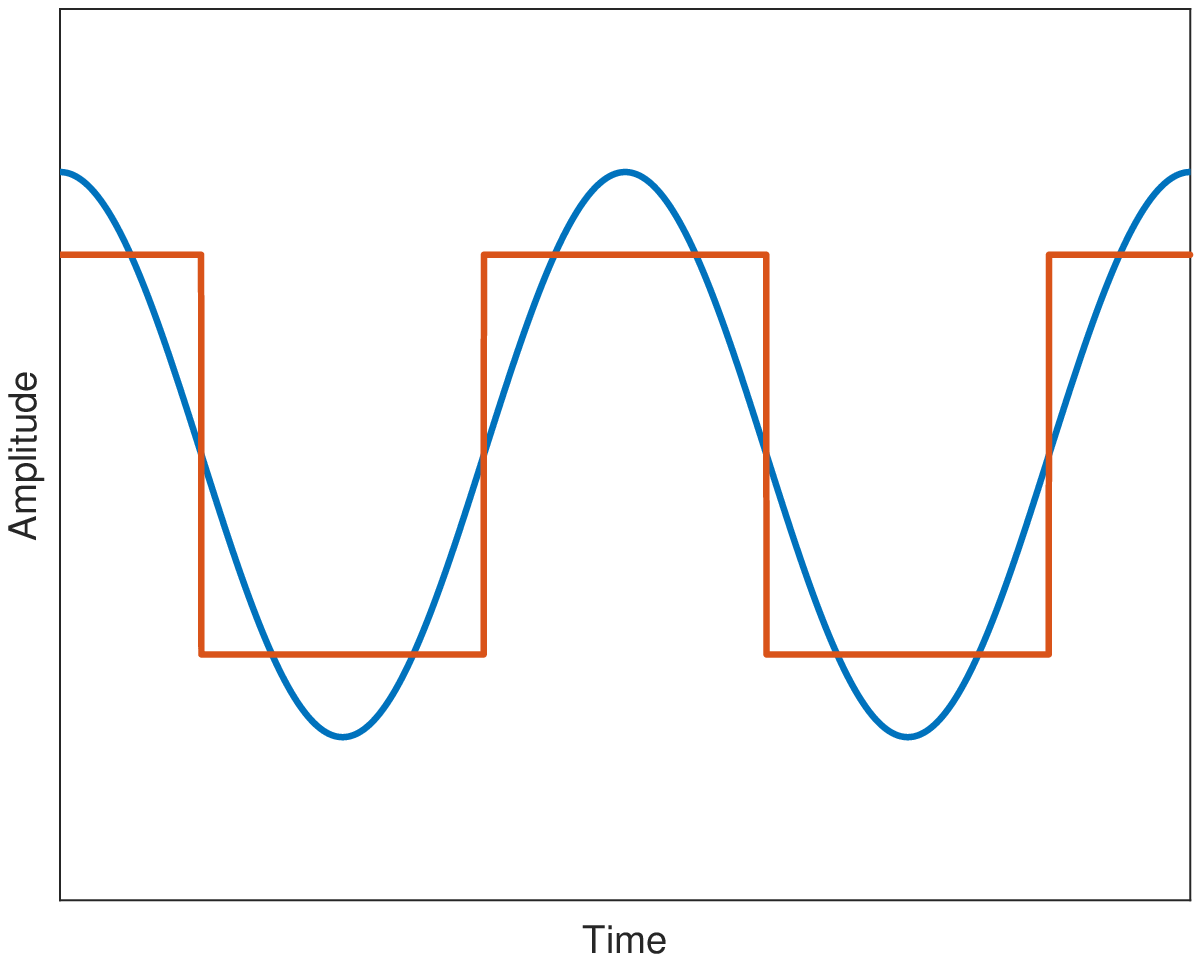}
			\includegraphics[width=0.31\linewidth]{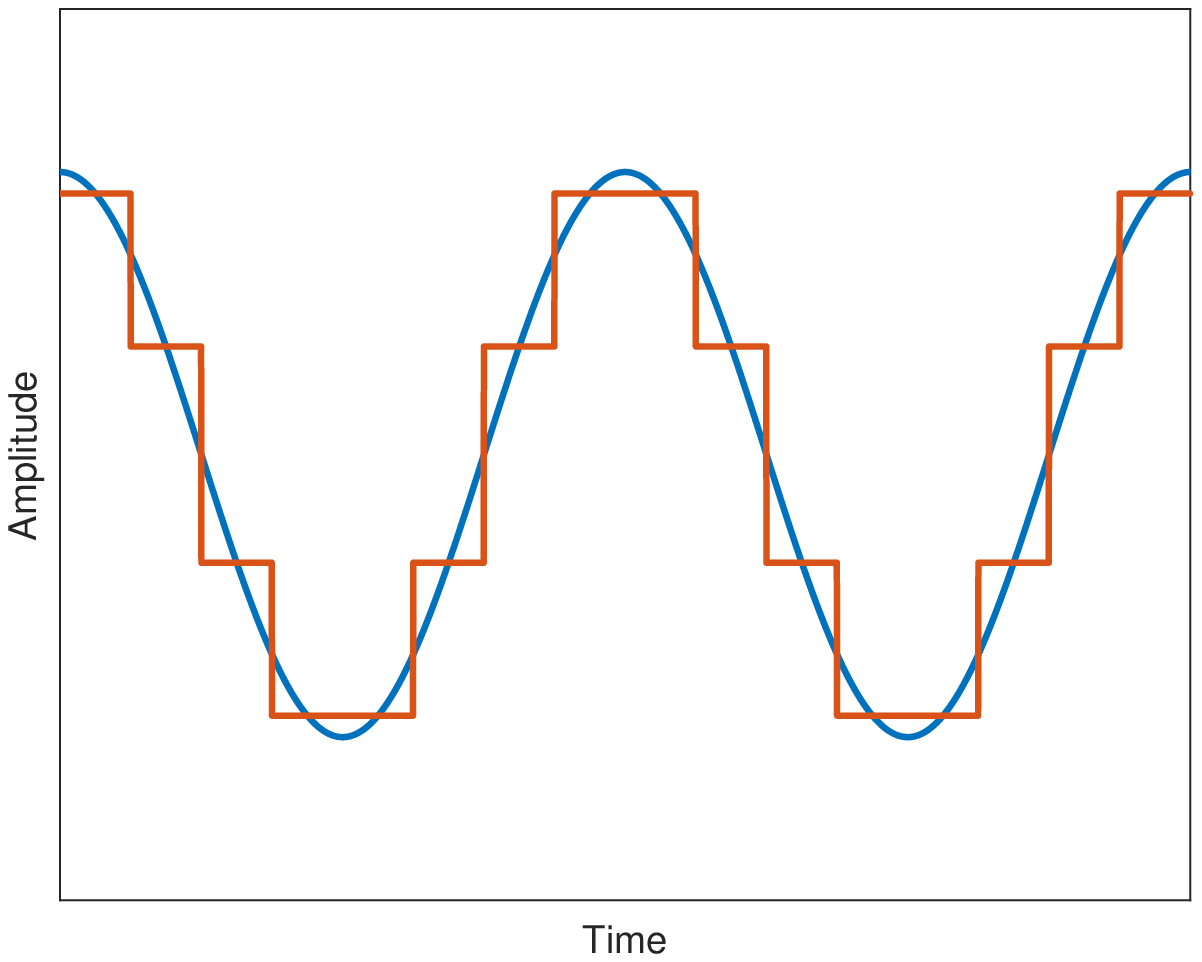}
		}
		\qquad
		\subfloat[]{
			\label{Fig:quad}
			\includegraphics[width=0.31\linewidth]{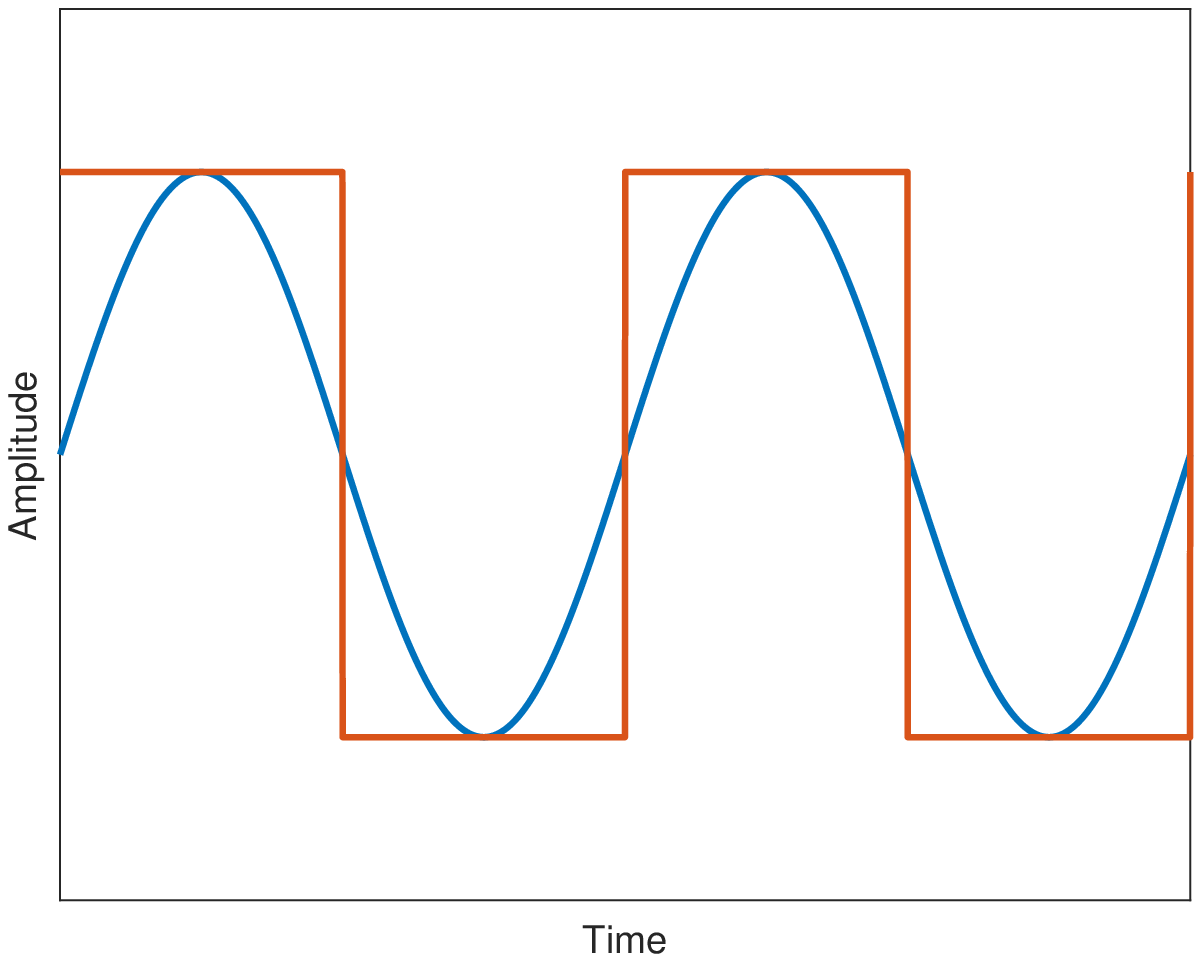}
			\includegraphics[width=0.31\linewidth]{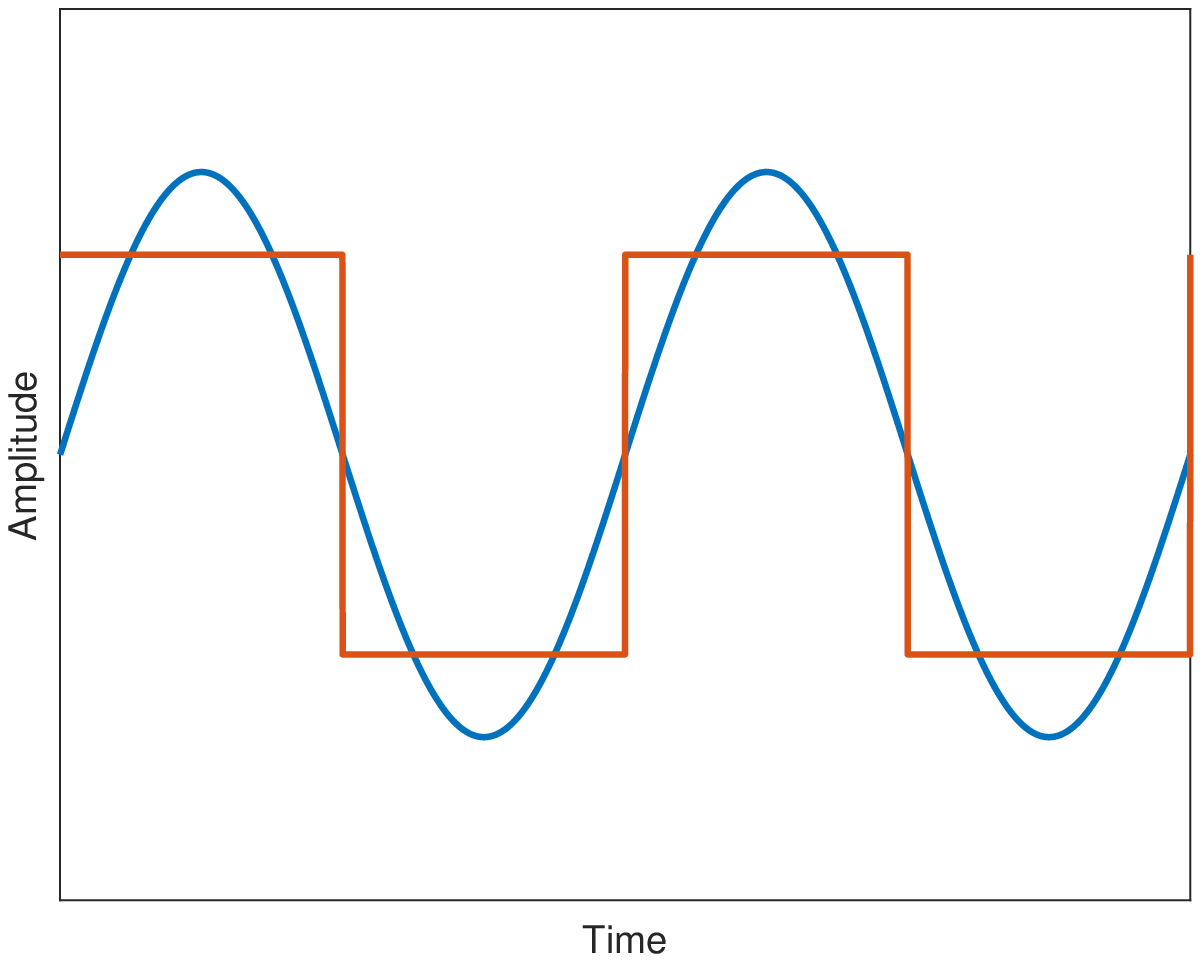}
			\includegraphics[width=0.31\linewidth]{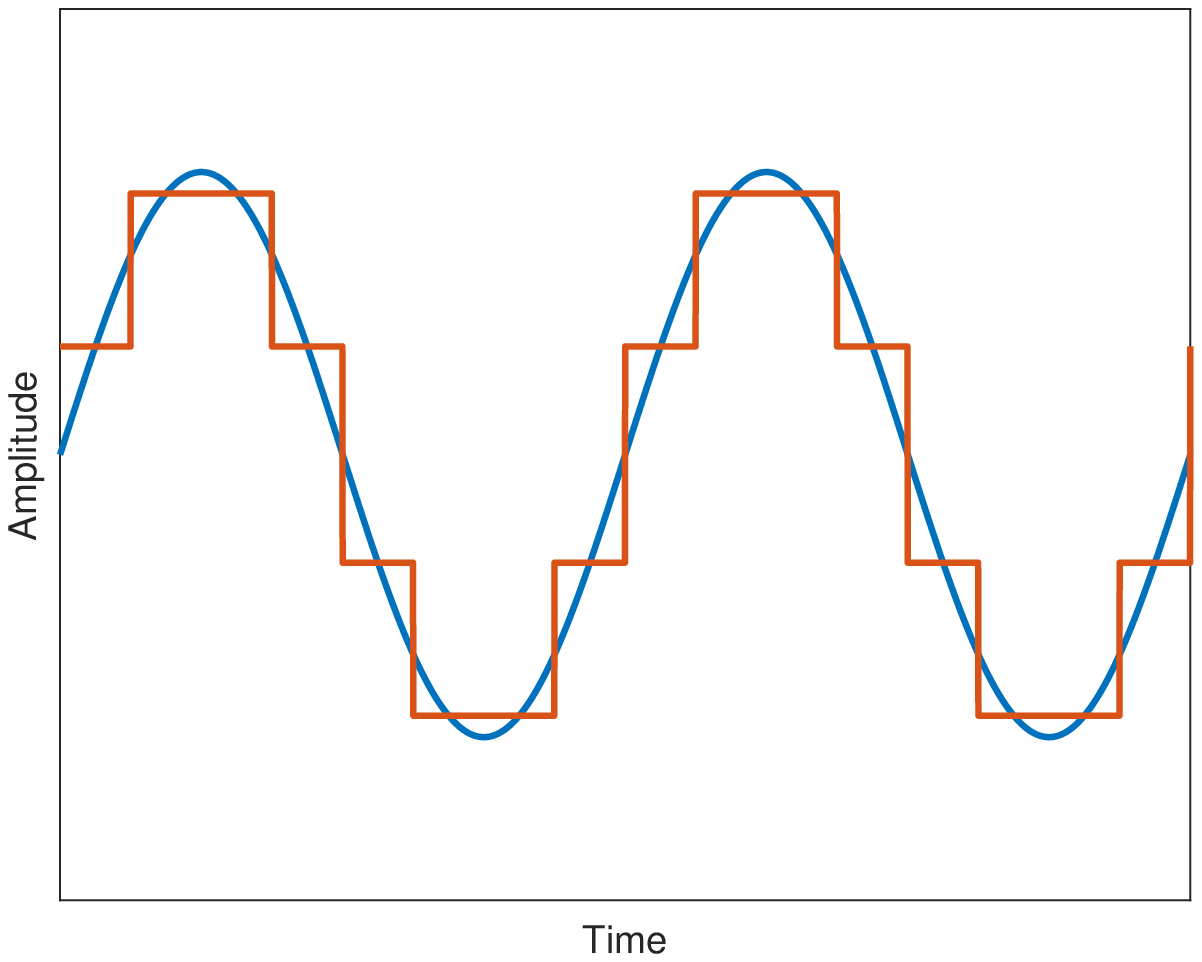}
		}
		\qquad
		\caption{IQ components comparison of backscattered waves (red) and sinusoidal waves (blue). The first, second, and third columns represent two, four, and eight phases, respectively: (a) phase diagram, (b) in-phase components, and (c) quadrature components.} 
		\label{Fig:IQcompare}
	\end{figure}
	


We first elaborate on some typical cases, i.e., using two, four, and eight phases to approximate sine and cosine. Then, we generalize the model. As shown in the first, second, and third columns in Fig.~\ref{Fig:IQcompare}, we compare the IQ components of approximated sinusoidal waves (marked with red) with pure sinusoidal waves (marked with blue). 

	If we consider only two phases for \ac{LB}, the in-phase component of the backscattered wave is zero while the quadrature component is a square wave. To avoid interference with the direct link, a frequency shift $\Delta f$ should be introduced. In normal non-\ac{BC} systems, the frequency shift is realized by multiplying the original signals centered at $f_c$ by a sinusoid with frequency $\Delta f$. However, due to the limited number of loads in \ac{BC} tags, it is introduced through multiplication with square waves with frequency  $\Delta f$, resulting in  additional harmonics \cite{selfsustainlora2021}. 
	More precisely, while moving the LoRa signals into the channel centered at $f_c+\Delta f$, a mirror copy centered at $f_c-\Delta f$ is also created. Also, there are harmonics centered at $f_c\pm 3\Delta f$, $f_c\pm 5\Delta f$, $f_c\pm 7\Delta f$, etc.
	
	If the system adopts four loads, both in-phase and quadrature components are square waves. This scheme cancels the mirror copies of the spectra. Nevertheless, the harmonics are preserved at $f_\textrm{c}-3\Delta f$, $f_\textrm{c}+5\Delta f$, $f_\textrm{c}-7\Delta f$, etc\cite{4pbsk}. When considering eight loads, two more voltage levels are added to the approximated sinusoids. The staircase-like waveforms cancels at least the harmonics centered at $f_\textrm{c}-3\Delta f$ and  $f_\textrm{c}+5\Delta f$\cite{3lorabsk}. Also, higher-order harmonics can be canceled by adding more phases if required.

	In this regard, unlike conventional LoRa, \ac{LB} has a finite number of phases in addition to several harmonics.	
	In the following, we  provide the signal expressions for the baseband signals accounting for the finite phases.

	\subsubsection{Continuous-Time Description}
	The baseband \ac{LB} signals are synthesized by switching between different loads so that they only have a limited number of phases, making them an approximation to the normal LoRa signals. 
	
	To generalize the model, let us consider the number of loads written in a form of $2^N$, where $N\in \offf{1,2,3,...}$. The phases are evenly distributed and rotationally symmetric in the phase diagram at position ${\left(2n-1\right)\pi}/{2^{N}},\,\forall n\in \left\{1,2,\cdots, 2^{N}\right\}$ with some examples shown in Fig.~\ref{Fig:phase_dia}.
	
	We introduce to the phase of conventional LoRa signal a mid-rise quantizer to perfectly model the discrete phases of \ac{LB}. The quantizer maps the infinite phases of the LoRa modulation into a finite set of phases in \ac{LB}. More precisely, the mid-rise quantization function with $2^{N}$ levels for input value $x\in\of{-\pi,\pi}$ is given by
	\begin{equation}
		\label{eq:midrisequantizer}
		\widetilde{Q}_N\left(x\right)=\dfrac{\lfloor 2^{N-1}x/\pi \rfloor +\frac{1}{2}}{2^{N-1}/\pi}, 
	\end{equation}
	where $ \lfloor \cdot \rfloor $ is the floor function\cite{harris1998handbook}. Let us also specify the periodicity of the quantization function as $\widetilde{Q}_N\of{x+2\pi} \triangleq \widetilde{Q}_N\of{x}$ for any real input value $x$, making it consistent with the periodic change of the phase. 
	
	 Let $T_\textrm{s}\triangleq M/B$ be the symbol duration, the instantaneous phase of LoRa modulation with symbol $a$ for interval $\off{0,T_\textrm{s}}$ can given in\cite{AEloramodulation} as
	\begin{equation}
		\hat{\phi}_a\of{t}=2\pi Bt\off{\frac{a}{M}-\frac{1}{2}+\frac{Bt}{2M}-u\of{t-\tau_a}}, 
	\end{equation}
	where $B$ denotes the bandwidth of LoRa, $M=2^{\text{SF}}$, $\text{SF}\in\offf{7,8,\cdots,12}$ denotes the spreading factor, and $\tau_a=({M-a})/{B}$ denotes the time instant of the sudden frequency change. Thus, the instantaneous phase of \ac{LB} with symbol $a$ is calculated as
	\begin{gather}
		\begin{aligned}
			\label{eq:LBphase}
			\phi_a\of{t}&= \widetilde{Q}_N\off{	\hat{\phi}_a\of{t}}.
		\end{aligned}		
	\end{gather}
	Thus, the complex envelope for \ac{LB} waveform of symbol $a$ with unit power is calculated using (\ref{eq:LBphase}) as
		\begin{align}
			x_a\of{t}&=\exp\offf{j\phi_a\of{t}}\\
			&=\exp\offf{j\widetilde{Q}_N\off{2\pi Bt\off{\frac{a}{M}-\frac{1}{2}+\frac{Bt}{2M}-u\of{t-\tau_a}}}}.\nonumber
		\end{align}
	
	

	\subsubsection{Discrete-Time Description}
	Let us consider the simple receiver implementation that sample the received signal at the chip rate, i.e., ${1}/{T_\text{c}}=M/{T_\text{s}}=B$, the $M$ samples in the interval  $\off{0,T_\textrm{s}}$, for $k\in\offf{0,1,\cdots,M-1}$, are
		\begin{align}
			\phi_a\of{kT_\text{c}}
			&=\widetilde{Q}_N\off{\frac{k\pi}{M}\of{2a-M+k}-2M\pi u\of{\frac{k+a-M}{B}}}\nonumber\\
			&=\widetilde{Q}_N\off{\frac{k\pi}{M}\of{2a-M+k}},
		\end{align}
	where the last equality is due to the fact that $2M\pi u\of{\cdot}$ is always an integer multiple of $2\pi$ for all $k$ and $\widetilde{Q}_N\of{x}$ is periodic of $2\pi$. 
	Normalizing the total energy of the samples, we have the discrete-time expression of the baseband \ac{LB} signal with $2^{N}$ quantization phases
		\begin{align}
		{x_a}\left[k\right]=\sqrt{\dfrac{1}{M}}\exp\left\{j \widetilde{Q}_N\left[\dfrac{k\pi}{M}\left(2a-M+k\right)\right]\right\}, \nonumber\\
	k\in\left\{0,1,\cdots,M-1\right\}.
		\end{align}

	\subsection{\acl{LB} Decoders}
	We study the error performance of \ac{LB} under two decoders, i.e., the \ac{ML} and \ac{FFT} decoders. We consider that the overall channel amplitude $\left|h\right|=\abs{h_1}\abs{h_2}$ is known to the \ac{Rx}, and it performs non-coherent detection.\footnote{{Non-coherent detectors for \ac{LB} can provide a good trade-off between the performance and complexity since the performance improvement using coherent detectors is only 0.7~dB  for traditional LoRa \cite{8835951}.}}
	
		\subsubsection{Maximum Likelihood Decoder}
	The optimum detector in an \ac{AWGN} channel is the \ac{ML} detector if all the symbols are equiprobable\cite{digitalCommunication2012}. The decoder performs cross-correlation on the received signal with all possible waveforms and chooses the symbol that has the largest absolute value of cross-correlation with the received signal, i.e.,
		\begin{align}
			\hat{a}&=\mathop{\arg\max}\limits_{0\leq i \leq M-1} \left| \left< \boldsymbol{r_a}, \boldsymbol{x^*_i}\right>\right|\\
			&=\mathop{\arg\max}\limits_{0\leq i \leq M-1}\abs{\sum_{k=0}^{M-1}{r_a}\left[k\right] \cdot {x_i^*}\left[k\right]},
		\end{align}
	where $ \hat{a} $ is the symbol decision made by the \ac{Rx}, $ {x_i^*}\left[k\right] $ is the complex conjugate of $ {x_i}\left[k\right]$.
The \ac{ML} decoder is the optimum decoder for \ac{LB} as shown in Appendix A. However, \ac{LB} is a relatively high-order modulation that reaches $\textrm{M}=4096$ for $\textrm{SF}=12$, resulting in a more complicated design and implementation of the decoder.

	\subsubsection{FFT Decoder}
	The process of FFT decoder follows three steps. First, The received signal is firstly multiplied by a complex down-chirp signal $ {x_\textrm{d}}\left[k\right]$
	\begin{equation}
		{\tilde{r}_a}\left[k\right]={r}_a\left[k\right]\, {x_\textrm{d}}\left[k\right]={r}_a\left[k\right]\, \sqrt{\frac{1}{M}}e^{-j2\pi \frac{k^2}{2M}+j\pi k}, 
	\end{equation}
	where the dechirped signal $ {\tilde{r}_a}\left[k\right] $ is called the twisted signal.
	The symbol decision is made by choosing the maximum value among M-bin \ac{dft} output on ${\tilde{r}_a}\left[k\right]$
		\begin{align}
			\hat{a}&=\mathop{\arg\max}\limits_{0\leq i \leq M-1}	\left|\textrm{DFT}\left({\tilde{r}_a}\left[k\right]\right)\right|\\
			&=\mathop{\arg\max}\limits_{0\leq i \leq M-1}\left|\sum_{k=0}^{M-1}{\tilde{r}_a}\left[k\right]\,  e^{-j2\pi ki/M}\right|. 
		\end{align}
	
		\begin{figure}[t]
		\psfrag{0.1}[c][c][0.5]{0.1}
		\psfrag{0.2}[c][c][0.5]{0.2}
		\psfrag{0.3}[c][c][0.5]{0.3}
		\psfrag{0.4}[c][c][0.5]{0.4}
		\psfrag{0.5}[c][c][0.5]{0.5}
		\psfrag{0.6}[c][c][0.5]{0.6}
		\psfrag{0.7}[c][c][0.5]{0.7}
		\psfrag{0.8}[c][c][0.5]{0.8}
		\psfrag{0.9}[c][c][0.5]{0.9}
		\psfrag{1}[c][c][0.5]{1}
		\centering
		\subfloat[\acl{LB} ($N=2$)]{
			\includegraphics[width=0.475\linewidth]{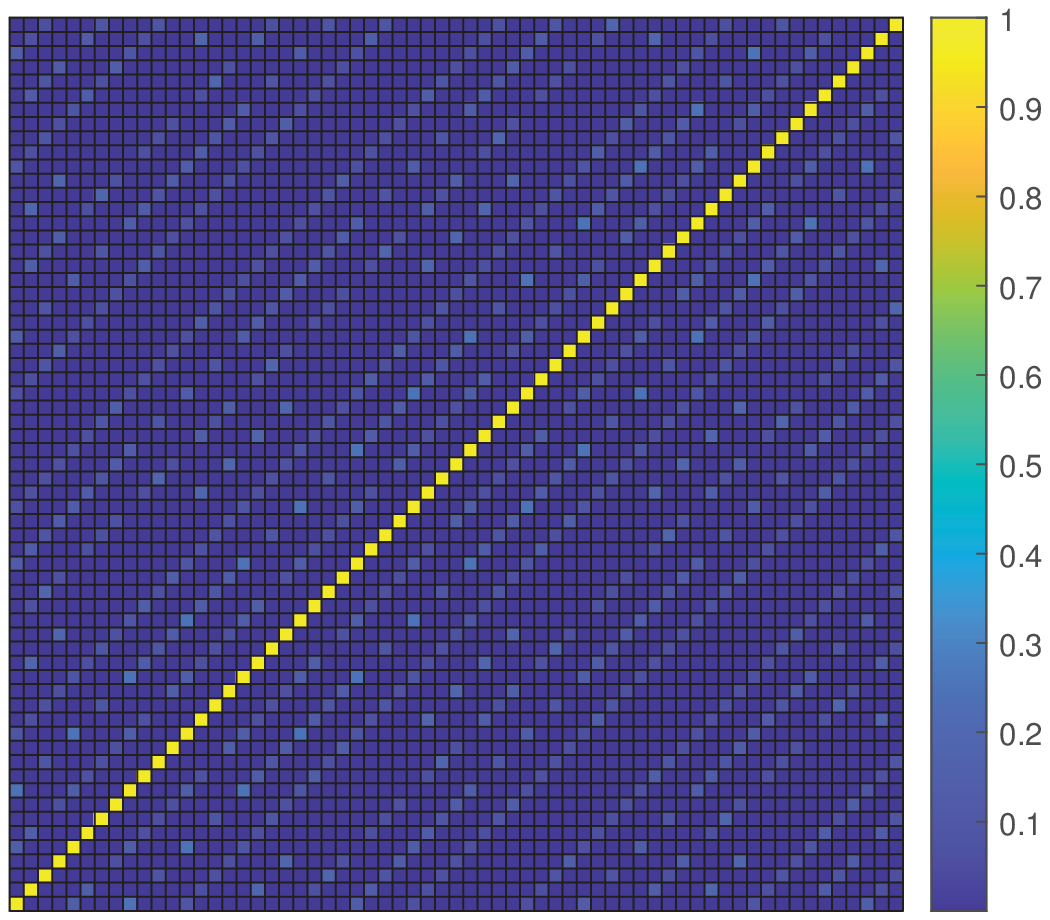}
			\label{fig:LBcorrelation}
		}
		\subfloat[Normal LoRa]{
			\includegraphics[width=0.475\linewidth]{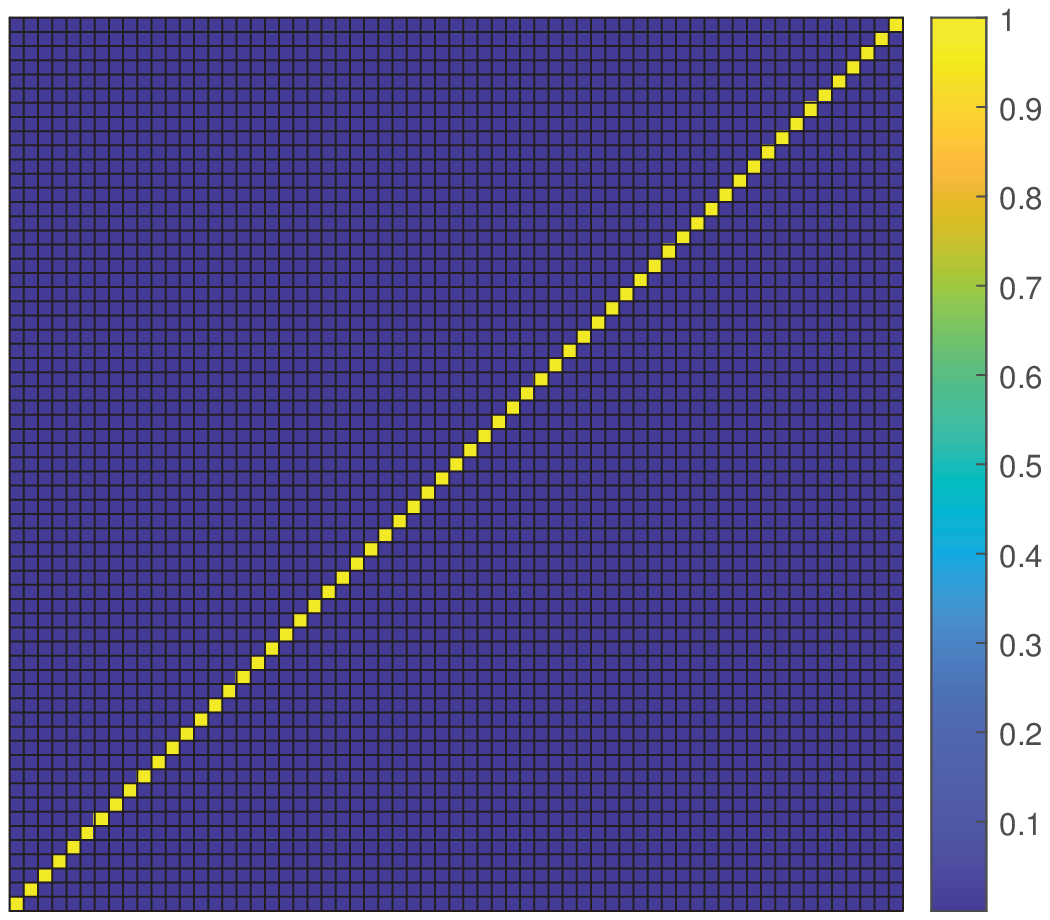}
			\label{fig:LoRacorrelation}
		}
		\caption{The cross-correlation of \ac{LB} and LoRa symbols when $\textrm{SF}=7$. }
		\label{fig:correlation}
	\end{figure}
	It has been proven that the two decoders are equivalent for norma LoRa\cite{AEloramodulation} but this is not valid for \ac{LB}. Also, the waveforms representing different symbols sampled at the chip rate $B$ are non-orthogonal. Fig.~\ref{fig:correlation} compares the cross-correlation of \ac{LB} waveforms (4 quantization phases) with that of normal LoRa at $\textrm{SF}=7$. It shows that the cross-correlation of \ac{LB} waveforms has numerous non-zero values while the cross-correlation between different symbols are all zero for normal LoRa\footnote{LoRa symbols are non-orthogonal, it can be proven orthogonal only when sampled at Nyquist rate.}. However, most of the non-zero correlation values are relatively small, indicating that the symbols are not strongly correlated.

{In Table.~\ref{tab:correlation_value_reduced}, we show the maximum  cross-correlation value between different \ac{LB} waveforms for $N\in \offf{2, 3, 4, 5}$ and $\text{SF}\in \offf{7,8, \cdots, 12}$. The cross-correlation generally decreases as $N$ and SF increase.} 

\begin{table}[t]
\caption{{Maximum  cross-correlation value between different \ac{LB} waveforms for $N\in \offf{2, 3, 4, 5}$ and $\text{SF}\in \offf{7,8, \cdots, 12}$.}}
\label{tab:correlation_value_reduced}
\resizebox{\columnwidth}{!}{%
\begin{tabular}{|c|c|c|c|c|c|c|}
\hline
    & SF=7  & SF=8  & SF=9  & SF=10 & SF=11 & SF=12 \\ \hline
N=2 & 0.250 & 0.156 & 0.156 & 0.117 & 0.086 & 0.067 \\ \hline
N=3 & 0.125 & 0.082 & 0.107 & 0.064 & 0.071 & 0.050 \\ \hline
N=4 & 0.000 & 0.000 & 0.053 & 0.032 & 0.043 & 0.024 \\ \hline
N=5 & 0.000 & 0.000 & 0.000 & 0.000 & 0.022 & 0.013 \\ \hline
\end{tabular}%
}

\end{table}

\section{Spectral analysis}
\label{sec:SpectralAnalysis}

The power spectrum of the \ac{LB} modulation is analytically determined in closed form in this section. We consider a source that produces a random symbol sequence where all the symbols are independent, identically distributed with \acl{r.v.}s $\mathcal{B}_n$. Assuming that the symbols are equiprobable, we have
\begin{equation}
P\of{\mathcal{B}_n=a}=\frac{1}{M}, \quad a\in\offf{0,1,\cdots,M-1}.
\end{equation}
The time sequence of the \ac{LB} waveforms can be expressed with a random process
\begin{equation}
	\label{eq:randomprocess}
 \mathcal{I}\of{t}=\sum_n x_s\of{t-nT_\text{s};\mathcal{B}_n}g_{T_\text{s}}\of{t-nT_\text{s}},
\end{equation}
where $g_{T_\text{s}}$ is a rectangular function with width ${T_\text{s}}$. The power spectrum density of the random process $ \mathcal{I}\of{t}$ can be derived as 
\begin{equation}
	\label{eq:total_psd}
	\mathcal{G}_ \mathcal{I}\of{f}=	\mathcal{G}_ \mathcal{I}^{\of{c}}\of{f}+\mathcal{G}_ \mathcal{I}^{\of{d}}\of{f}.
\end{equation}
where $\mathcal{G}_ \mathcal{I}^{\of{c}}\of{f}$ and $\mathcal{G}_ \mathcal{I}^{\of{d}}\of{f}$ are the continuous and a discrete parts of the spectrum, respectively. The continuous and discrete spectrum in (\ref{eq:total_psd}) can be obtained by applying frequency domain analysis of randomly modulated signals for the random process \cite{benedetto1999principles}
\begin{align}
	\mathcal{G}_ \mathcal{I}^{\of{c}}\of{f}&=\frac{1}{T_\text{s}}\off{\sum_{a=0}^{M-1}\frac{1}{M}\abs{S_a\of{f}}^2-\abs{\sum_{a=0}^{M-1}\frac{1}{M}S_a\of{f}}^2},\label{eq:GIC}\\
	\mathcal{G}_ \mathcal{I}^{\of{d}}\of{f}
	&=\frac{1}{M^2T^2_\text{s}}\sum_{l=-\infty}^{\infty}\abs{\sum_{a=0}^{M-1}S_a\of{\frac{lB}{M}}}^2\delta\of{f-\frac{lB}{M}}\label{eq:GID},
\end{align}
where $\offf{S_a\of{f}}_{a=0}^{M-1}$ are the Fourier transforms of the waveforms $\offf{x_a\of{t}}_{a=0}^{M-1}$ from the modulator. 

The Fourier transform of the complex envelope can be expressed as
\begin{align}
		S_a\of{f}&=\int_{0}^{T_\text{s}}e^{j\widetilde{Q}_N\off{2\pi Bt\off{\frac{a}{M}-\frac{1}{2}+\frac{Bt}{2M}-u\of{t-\frac{M-a}{B}}}}}e^{-j2\pi ft}\mathrm{d}t\nonumber\\
		&=\int_{0}^{\frac{M-a}{B}}e^{j \widetilde{Q}_N\off{2\pi Bt\of{\frac{a}{M}-\frac{1}{2}+\frac{Bt}{2M}}}}e^{-j2\pi f t}\mathrm{d}t+\nonumber\\
	&\quad \quad \int_{\frac{M-a}{B}}^{\frac{B}{M}}e^{j \widetilde{Q}_N\off{2\pi Bt\of{\frac{a}{M}-\frac{3}{2}+\frac{Bt}{2M}}}}e^{-j2\pi f t}\mathrm{d}t.\label{eq:Sa}
	\end{align}

Leveraging the periodicity of $\widetilde{Q}_N\of{\cdot}$, we rewrite the second term in (\ref{eq:Sa}) as
\begin{equation}
\int_{\frac{M-a}{B}}^{\frac{B}{M}}e^{j\pi \widetilde{Q}_N\off{2\pi Bt\of{\frac{a}{M}-\frac{3}{2}+\frac{Bt}{2M}}+2\pi\of{M-a}}}e^{-j2\pi f t}\mathrm{d}t.
\end{equation}
Let us also define $f_1\of{t}\triangleq 2\pi Bt\of{\frac{a}{M}-\frac{1}{2}+\frac{Bt}{2M}}$, and  $f_2\of{t}\triangleq 2\pi Bt\of{\frac{a}{M}-\frac{3}{2}+\frac{Bt}{2M}}+2\pi \of{M-a}$ as the two terms within $\widetilde{Q}_N\of{\cdot}$ of two integral parts. Let us define $f\of{t}$ as
\begin{equation}\nonumber
	f\of{t}\triangleq	\left\{
	\begin{aligned}
		f_1\of{t}\quad&,\quad \quad \ 0\quad\leq t<\frac{M-a}{B}\\
		f_2\of{t}\quad&,\quad\frac{M-a}{B}\leq t\leq\quad\frac{M}{B}
	\end{aligned}
	\right.
\end{equation}
It is worth noting that, $f\of{t}$ is a continuous function with $f_1\of{\frac{M-a}{B}}=f_2\of{\frac{M-a}{B}}$. 
Thus, the sum of two separated integrals in (\ref{eq:Sa}) can be replaced with one integral
\begin{equation}\label{eq:Sa1}
	S_a\of{f}=\int_{0}^{\frac{B}{M}}e^{j \widetilde{Q}_N\off{f\of{t}}}e^{-j2\pi f t}\mathrm{d}t.
\end{equation}

\begin{figure}[t]
	\centering
	\pgfplotsset{every axis/.append style={
		font=\footnotesize,
		line width=1pt,
		legend style={font=\footnotesize, at={(0.99,0.99)}},legend cell align=left,nodes={scale=0.8, transform shape}},
} %
\pgfplotsset{compat=1.13}
	\begin{tikzpicture}
\begin{axis}[
xlabel near ticks,
ylabel near ticks,
grid=both,
xlabel={$t$},
ylabel={$\widetilde{Q}_N\off{f\of{t}}$},
	ytick={-1, -0.5,0,0.5,1},
	yticklabels={$-\pi$,$-\frac{\pi}{2}$,$0$,$\frac{\pi}{2}$,$\pi$},
	xtick={ 0,0.76393202,2,	4,	5.2360680,6,6.6055513,7.3944487,8},
	xticklabels={$t_0$,$t_1$,$t_2$,$t_3$,$t_4$,$t_5$,$t_6$,$t_7$,$t_\psi$},
yticklabel style={/pgf/number format},
width=0.9\linewidth,
	xmin= 0, xmax=8,
	ymin=-1, ymax=1,
ylabel style={font=\large},
xlabel style={font=\Large},
]
\addplot[blue!70!white]table {Figures/PSD/SaExample/SaExampleData.dat};
\end{axis}
\end{tikzpicture}
	\caption{An example of $\widetilde{Q}_N\off{f\of{t}}$ as a function of time at $N=2$.}
	\label{fig:saexample}
\end{figure}
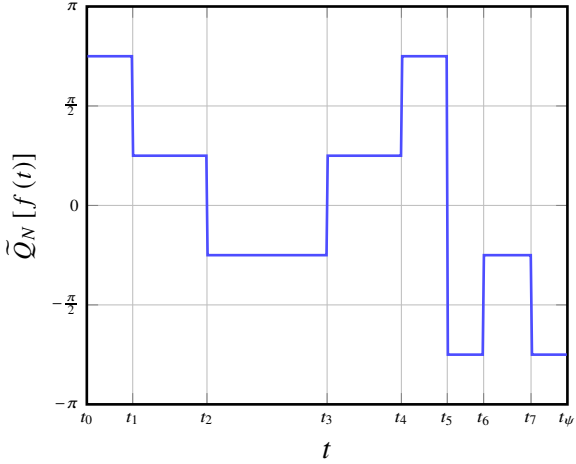

It is challenging to directly compute the integral in \eqref{eq:Sa1} with the quantization function inside. However, one possible solution is to divide the integral interval into slots, as shown in Fig.~\ref{fig:saexample}. The value of $\widetilde{Q}_N\off{f\of{t}}$ remains unchanged in each intervals divided by a time set $\offf{t_m}_{m=1}^{\psi}$. Details about obtaining $\offf{t_m}_{m=1}^{\psi}$ are provided in Appendix B.
The integral in each slot can be calculated as
	\begin{align}
		I_m\of{f}&=\int_{t_m}^{t_{m+1}}e^{j \widetilde{Q}_n\off{f\of{t}}}e^{-j2\pi f t}\mathrm{d}t\\
		&=\frac{j}{2 f}e^{j\pi \widetilde{Q}_n\off{f\of{\frac{t_m+t_{m+1}}{2}}}}\of{e^{-j2\pi f t_{m+1}}-e^{-j2\pi f t_m}},\nonumber
	\end{align}
where $({t_m+t_{m+1}})/{2}$ can be replaced by any real value between $t_m$ and $t_{m+1}$. 

Since we have $f\of{0}=f\of{{M}/{B}}=0$, the overall integral interval begins with $t_0$ and ends with $t_\psi$. Thus, the integral in (\ref{eq:Sa}) is calculated as a sum of $I_m\of{f}$
\begin{gather}
\label{eq:safinal}
	\begin{aligned}
		S_a\of{f}=\frac{j}{2\pi f}\sum_{m=1}^{\psi-1}e^{j\pi \widetilde{Q}_N\off{f\of{\frac{t_m+t_{m+1}}{2}}}}\of{e^{-j2\pi f t_{m+1}}-e^{-j2\pi f t_m}}.
	\end{aligned}
\end{gather} 
By substituting (\ref{eq:safinal}) into (\ref{eq:GIC}) and (\ref{eq:GID}), we can calculate both the continuous and discrete parts of the spectrum for baseband \ac{LB} signals.


\section{Performance Analysis}
\label{sec:performance}
In this section, we derive the \ac{SER} of the \ac{LB} communication system in terms of different decoders, channel models, and power strategies.
\subsection{\acl{LB} \ac{SER} Performance over AWGN Channels }
The \ac{SER} in an \ac{AWGN} channel is a conditional probability conditioned on the overall channel gain $h$. Thus, we consider $h$ as a constant in this subsection. We define the \ac{r.v.} $\mathcal{L}^{D}_{\left(a,i\right)}$ as the absolute value of $i$-th bin of $D\in \mathbb{D}\triangleq \offf{\textrm{ML},\textrm{FFT}}$ decoder when symbol $a$ is transmitted. For \ac{ML} decoder, $\mathcal{L}^{\textrm{ML}}_{\left(a,i\right)}$ is calculated as
	\begin{align}
		\mathcal{L}^{\text{ML}}_{\left(a,i\right)}&=\abs{\sum_{k=0}^{M-1}{r_a}\left[k\right] {x_i^*}\left[k\right] }\\
		&=
		\abs{h\sqrt{E_\text{s}}
		{\xi_{\of{a,i}}^{\text{ML}}}
		+{\mathcal{W}^{\text{ML}}_{\left(a,i\right)}}},\label{eq:LaiML}
	\end{align}
where $\xi_{\of{a,i}}^{\text{ML}}=\sum_{k=0}^{M-1} {x_a}\left[k\right] {x_i^*}\left[k\right] $ is the cross-correlation between the transmitted waveform and the $i$-th reference signal and $\mathcal{W}^{\text{\text{ML}}}_{\left(a,i\right)}=\sum_{k=0}^{M-1}{\omega}\off{k}{x_i^*}\left[k\right]$ is the noise projection of the reference signals, following complex Gaussian distribution with variance $2\sigma^2$. For \ac{FFT} decoder, $\mathcal{L}^{\textrm{FFT}}_{\left(a,i\right)}$ is 
	\begin{align}
		\mathcal{L}^{\text{FFT}}_{\left(a,i\right)}&=\abs{\textrm{DFT}\left( {r_a}\left[k\right] {x_\textrm{d}}\left[k\right] \right) }\\
		&=\abs{
		h\sqrt{E_\text{s}}
		{\xi_{\of{a,i}}^{\text{FFT}}}
		+{\mathcal{W}^{\text{FFT}}_{\left(a,i\right)}}},\label{eq:LaiFFT}
	\end{align}
where $\xi_{\of{a,i}}^{\text{FFT}}=\textrm{DFT}\left( {x_a}\left[k\right] {x_\textrm{d}}\left[k\right] \right) $ denotes the i-th bin of the \ac{dft} for the dechirped reference signals and $\mathcal{W}^{\text{FFT}}_{\left(a,i\right)}=\text{DFT}\of{{\omega}\off{k}{x_\textrm{d}}\left[k\right]}$ is the $i$-th bin of the \ac{dft} for the dechirped complex white Gaussian noise, which also follows complex Gaussian distribution with variance $2\sigma^2$. 
Since (\ref{eq:LaiML}) and (\ref{eq:LaiFFT}) have the same pattern, i.e., the absolute value of a sum of a complex constant and a complex Gaussian \ac{r.v.}, \begin{equation}
	\mathcal{L}^{D}_{\left(a,i\right)}=\abs{
		h\sqrt{E_\textrm{s}}\xi^{D}_{\left(a,i\right)}+\mathcal{W}^{D}_{\left(a,i\right)}
	},
\end{equation}
which follows Rician distribution with shape parameter $\kappa^D_{\left(a,i\right)}={\left|h\sqrt{E_\textrm{s}}\xi_{\of{a,i}}^D\right|^2}/{2\sigma^2}$, as shown in Appendix C. Without losing generality, if $\xi_{\of{a,i}}^D=0$, the shape parameter is zero so that the Rician distribution becomes Rayleigh distribution.

It is worth noting that the \acp{r.v.} in the set $\offf{\mathcal{L}^{\text{ML}}_{\left(a,i\right)}}_{i=0}^{M-1}$  are not independent because $\mathcal{W}^{\text{ML}}_{\left(a,i\right)}$ are the noise projection of non-orthogonal waveforms, as shown in Fig.~\ref{fig:LBcorrelation}. 
On the other hand, the \acp{r.v.} in the set $\offf{\mathcal{L}^{\text{FFT}}_{\left(a,i\right)}}$ are independent since the noise is projected to the directions of orthogonal basis in the \ac{dft} process. The dependence of $\offf{\mathcal{L}^{\text{ML}}_{\left(a,i\right)}}_{i=0}^{M-1}$ make it challenging to find their joint distribution. However, they are approximately independent for the following three reasons. 
First, as shown in Fig.~\ref{fig:LBcorrelation}, most of the cross-correlation values are zero, namely, most of the \acp{r.v.} within $\offf{\mathcal{L}^{\text{ML}}_{\left(a,i\right)}}_{i=0}^{M-1}$ are independent. 
Second, the non-zero cross-correlation values are relatively small so that the dependent \acp{r.v.} are not strongly correlated. Also, the non-zero cross-correlation values decrease as the number of quantization phases or the spreading factor increases. 
Nevertheless, ignoring the correlation makes the derived expression an upper bound to the exact \ac{SER}\cite[Section~4.5-4]{digitalCommunication2012}. In the following, we assuming that $\offf{\mathcal{L}^{\text{ML}}_{\left(a,i\right)}}_{i=0}^{M-1}$ are independent so that the derivation applies to both decoders.

To obtain an expression for the \ac{SER}, let us  define the instantaneous \ac{SNR} as
\begin{equation}
	\gamma=\dfrac{\abs{h}^2E_\textrm{s}/T_\textrm{s}}{2\sigma^2B}=\dfrac{\abs{h}^2E_\textrm{s}}{2\sigma^2M}.
\end{equation}
The \ac{PDF} and \ac{CDF} of $ \mathcal{L}^{D}_{\left(a,i\right)}$  conditioned on the overall channel gain $h$ can be written for $D\in \mathbb{D}$ as
\begin{align}
	f_{\mathcal{L}^{D}_{\left(a,i\right)}|h}\left(l\right)&=
	\frac{l}{\sigma^2}\exp \left[\dfrac{-\left(l^2+C^2\right)}{2\sigma^2}\right]
	 I_0\left(\dfrac{lC}{\sigma^2}\right), \label{eq:ldaipdf} \\
	F_{\mathcal{L}^{D}_{\left(a,i\right)}|h}\left(l\right)&=1-Q_1\left(\dfrac{C}{\sigma}, \dfrac{l}{\sigma}\right),\label{eq:ldaicdf}
\end{align}
where $C=\abs{h\sqrt{E_\textrm{s}}\xi_{\of{a,i}}^D}$, $ Q_1\left(\cdot\right) $ is the Marcum Q-function of order one \cite[eq: 7]{705532},  $I_0\left(\cdot\right)$ is the modified Bessel function of the first kind and order zero\cite[eq: 9.6.1]{1964handbook}. 

The detection error happens when $\hat{a}\neq a$, i.e., $\mathcal{L}^D_{\left(a,a\right)}$ is not maximum among the set $\offf{\mathcal{L}^{D}_{\left(a,i\right)}}_{i=0}^{M-1}$. In other words, the \ac{SEP} can be calculated by comparing the correct bin, namely,  $\mathcal{L}^D_{\left(a,a\right)}$, with the maximum of the noisy bins. Thus, 
we define $\hat{\mathcal{L}}^{D}_{\left(a\right)}\triangleq \mathop{\max}\limits_{ i,i\neq a}\offf{\mathcal{L}^D_{\left(a,i\right)}}$ whose 
 \ac{CDF} of $\hat{\mathcal{L}}^{D}_{\left(a\right)}$ can be computed using order statistics as
	\begin{align}
		F_{\hat{\mathcal{L}}^{D}_{\left(a\right)}|h}\left(l\right)&=\prod_{i=0,i\neq a}^{M-1}
		F_{\mathcal{L}^{D}_{\left(a,i\right)}|h}\left(l\right)\nonumber\\
		&=\prod_{i=0,i\neq a}^{M-1}
		\left[1-Q_1\left(\dfrac{\left|h\sqrt{E_\textrm{s}}\xi^D_{\left(a,i\right)}\right|}{\sigma}, \dfrac{l}{\sigma}\right)\right]. \label{eq:lhatcdf}
	\end{align}
The conditional \ac{SEP} using $D\in \mathbb{D}$ decoder given a transmit symbol $a$ is calculated as
	\begin{align}
		P^D_{\textrm{e}|a}&=P\left(\hat{\mathcal{L}}^{D}_{\left(a\right)}>\mathcal{L}^{D}_{\left(a,a\right)}\big| h\right) \nonumber\\
		&=\int_{0}^{\infty}\left[1-F_{\hat{\mathcal{L}}^D_{\left(a\right)}|h}\left(l\right)\right]\, f_{\mathcal{L}^D_{\left(a,a\right)}|h}\left(l\right)\mathrm{d}l,	\label{eq:peaint}
	\end{align}
where the shape parameter of  $\mathcal{L}^D_{\left(a,a\right)}$ is
    \begin{align}
    \kappa^D_{\left(a,a\right)}&=\left|h\sqrt{E_\text{s}}\xi^D_{\left(a,a\right)}\right|^2/2\sigma^2\\
    &= \abs{\xi^D_{\left(a,a\right)}}^2 \, M\,\gamma.    \label{eq:kappaaa}
    \end{align}
The Rician distributed \ac{r.v.} $\mathcal{L}^{D}_{\left(a,a\right)}$ can be approximated as a Gaussian \ac{r.v.} if the shape parameter $\kappa^D_{\left(a,a\right)}$ is larger than 2 \cite{gudbjartsson1995rician}. 
In our setting, as shown in (\ref{eq:kappaaa}),  $\xi^D_{\left(a,a\right)}\approx 1$ and $M=2^{\textrm{SF}}$ ranging from $2^7=128$ to $2^{12}=4096$ is a large number so that  $\mathcal{L}^{D}_{\left(a,a\right)}$ can be approximated as a Gaussian \ac{r.v.} in an acceptable \ac{SNR} range. For example, if we consider $\textrm{SF}=8$ and \ac{ML} decoder, we have $\xi^{\textrm{ML}}_{\left(a,a\right)}=1$ and $M=256$. The proper \ac{SNR} range for the approximation is $\gamma\geq 1/128 \approx -21.1~\textrm{dB}$, which covers most of the \ac{SER} range of interest.
Hence, we have
\begin{equation}
	f_{\mathcal{L}^D_{\left(a,a\right)}|h}\left(l\right)\approx \dfrac{1}{\sqrt{2\pi\sigma_a^2}} \exp\left[-\frac{\left(l-\mu_a\right)^2}{2\sigma_a^2}\right],
\end{equation}
where $\mu_a$ and $\sigma_a$ are the mean and variance of $\mathcal{L}^D_{\left(a,a\right)}$, respectively, i.e., 
\begin{align}
	\mu_a&=\mathbb{E}\off{\mathcal{L}^D_{\left(a,a\right)}}=
	\sigma\sqrt{\pi/2}\cdot L_{1/2}\of{-\kappa^D_{\left(a,a\right)}},
	\\
	\sigma_a^2&=\mathbb{V}\off{\mathcal{L}^{D}_{\left(a,a\right)}}=
	2\sigma^2\of{1+\kappa^D_{\left(a,a\right)}}-\mu_a^2, 
\end{align}
where $ L_{q}\left(\cdot\right) $ denotes a Laguerre polynomial\cite[eq: 13.6.27]{1964handbook}, and for the case $q=1/2$, we have
\begin{equation}
	L_{1/2}\left(x\right)=e^{x/2}\left[\left(1-x\right)I_0\left(-\dfrac{x}{2}\right)-xI_1\left(-\dfrac{x}{2}\right)\right] . 
\end{equation}
\begin{table}[t]
\caption{{Typical values of $\kappa^D_{\of{a,a}}$, $\mu_a$, and $\sigma_a^2$ for $N \in\offf{2,3,4,5}$ at $\text{SF}=7$, $\gamma=-10~\textrm{dB}$, and $a=h=E_s=1$.}}
\label{tab:typical values}
\resizebox{\columnwidth}{!}{
\begin{tabular}{|c|cc|cc|cc|}
\hline
\multirow{2}{*}{$N$} & \multicolumn{2}{c|}{$\kappa^D_{\of{a,a}}$} & \multicolumn{2}{c|}{$\mu_a$} & \multicolumn{2}{c|}{$\sigma_a^2$} \\ \cline{2-7} 
  & \multicolumn{1}{c|}{ML}    & FFT   & \multicolumn{1}{c|}{ML}    & FFT   & \multicolumn{1}{c|}{ML}    & FFT   \\ \hline
2 & \multicolumn{1}{c|}{12.80} & 10.51 & \multicolumn{1}{c|}{26.11} & 23.75 & \multicolumn{1}{c|}{25.08} & 24.96 \\ \hline
3 & \multicolumn{1}{c|}{12.80} & 12.27 & \multicolumn{1}{c|}{26.11} & 25.58 & \multicolumn{1}{c|}{25.08} & 25.05 \\ \hline
4 & \multicolumn{1}{c|}{12.80} & 12.71 & \multicolumn{1}{c|}{26.11} & 26.01 & \multicolumn{1}{c|}{25.08} & 25.07 \\ \hline
5 & \multicolumn{1}{c|}{12.80} & 12.78 & \multicolumn{1}{c|}{26.11} & 26.09 & \multicolumn{1}{c|}{25.08} & 25.08 \\ \hline
\end{tabular}
}
\end{table}
{In Table.~\ref{tab:typical values}, we show some typical values of $\kappa^D_{\of{a,a}}$, $\mu_a$, and $\sigma_a^2$ as a function of $N$ for both \ac{ML} and \ac{FFT} decoders. }

The complexity of evaluating integral (\ref{eq:peaint}) mainly comes from the product of $M-1$ non-identical Rician \ac{CDF}s.  As a possible solution, we can leverage an asymptotic approximation of the order statistics of a sequence of independent and non-identically distributed (i.n.i.d.) Rician \acp{r.v.}  \cite{9915518} using the \ac{EVT}. Nevertheless, it shall result in an  exponential-in-exponential expression, rendering difficulties in the  integration. 

Alternatively, we consider  Gauss-Hermite quadrature to evaluate the integral numerically. In this procedure, the integral is converted to a weighted sum of function values  using the Gauss-Hermite quadrature \cite{1964handbook}. Since the Rician \ac{PDF} (\ref{eq:ldaipdf}) is zero for $l<0$, the lower limit of integral (\ref{eq:peaint}) can be substituted with $-\infty$. Thus, we get
\begin{gather}
	\begin{aligned}
		P^D_{\textrm{e}|a}&\overset{}{\approx}\int_{-\infty}^{\infty} \dfrac{\left[1-F_{\hat{\mathcal{L}}^D_{\left(a\right)}|h}\left(l\right)\right]}{\sqrt{2\pi\sigma_a^2}}\, \exp\left[-\frac{\left(l-\mu_a\right)^2}{2\sigma_a^2}\right] \mathrm{d}l.
	\end{aligned}
\end{gather}
Let us define $\hat{l}\triangleq\of{l-\mu_a}{\sqrt{2}\sigma_a}$ and substitute $l$ with $\sqrt{2}\sigma_a\hat{l}+\mu_a$ so that the integral becomes
	\begin{align}
		P^D_{\textrm{e}|a}&\overset{}{\approx}\int_{-\infty}^{\infty} {\left[1-F_{\hat{\mathcal{L}}^D_{\left(a\right)}|h}\left(\sqrt{2}\sigma_a\hat{l}+\mu_a\right)\right]} \frac{\exp\off{-\hat{l}^2}}{\sqrt{\pi}} \mathrm{d}\hat{l}\\
		&\overset{}{\approx} \dfrac{1}{\sqrt{\pi}}\sum_{t=1}^{N_\textrm{GH}}\omega_{t} \left[1-F_{\hat{\mathcal{L}}^D_{\left(a\right)}|h}\left(\sqrt{2}\sigma_a x_{t}+\mu_a\right)\right], 	\label{eq:GHquad}
	\end{align}
where  $ N_\textrm{GH} $ is the number of function samples used to approximate the integral, $x_{t}$ are the roots of the physicists' version of the Hermite polynomial $ H_{N_\textrm{GH}}\left(x\right) $, and  $\omega_{t} $ are the corresponding weights. The integration formula and corresponding weights  can be found in \cite[eq: 25.4.46]{1964handbook}.


We assume that the transmit symbols are equiprobable
so that the average \ac{SER} of \ac{LB} in the \ac{AWGN} channel can be expressed as

	\begin{align}
		\overline{P^D_{\textrm{e}|h}}&=\frac{1}{M}\sum_{a=0}^{M-1} P^D_{\textrm{e}|a} \nonumber\\
		&
		=\frac{1}{M\sqrt{\pi}}\sum_{a=0}^{M-1}\sum_{t=1}^{N_\textrm{GH}}\omega_{t} \left[1-F_{\hat{\mathcal{L}}^D_{\left(a\right)}|h}\left(\sqrt{2}\sigma_a x_{t}+\mu_a\right)\right].	\label{eq:avgSERAWGN}
	\end{align}

	\subsection{\acl{LB} \ac{SER} Performance over Double Nakagami-m Fading Channels}
	We consider a double Nakagami-m fading scenario with different shape parameters $\offf{m_1, m_2}$ and spread parameters $\offf{\Omega_1, \Omega_2}$. The spread parameters are related to the link distance, i.e., $\Omega_1\propto 1/d_1^{2}$ and $\Omega_2\propto 1/d_2^{2}$, where $\offf{d_1, d_2}$ are the distances of the two links. Let us define \acp{r.v.} $\abs{\mathcal{H}_1}$ and $\abs{\mathcal{H}_2}$ as the channel amplitudes of \ac{Tx}-tag and tag-\ac{Rx} links, respectively. The \ac{PDF}s of $\abs{\mathcal{H}_1}$ and $\abs{\mathcal{H}_2}$ can be expressed as
	\begin{align}
		f_{\abs{\mathcal{H}_1}}\left(\abs{h_1}\right)&=\dfrac{2m_1^{m_1}}{\Gamma\left(m_1\right)\Omega_1^{m_1}}|h_1|^{2m_1-1}\exp\off{-\dfrac{m_1}{\Omega_1}|h_1|^2}, \\
		f_{\abs{\mathcal{H}_2}}\left(\abs{h_2}\right)&=\dfrac{2m_2^{m_2}}{\Gamma\left(m_2\right)\Omega_2^{m_2}}|h_2|^{2m_2-1}\exp\off{-\dfrac{m_2}{\Omega_1}|h_2|^2}. 
	\end{align}
	Let us also define the \ac{r.v.} $\abs{\mathcal{H}} \triangleq \abs{\mathcal{H}_1}\abs{\mathcal{H}_2}$ as the overall channel amplitude, whose \ac{PDF} is governed by \cite{nNakagami2007n}
	\begin{equation}
		\label{eq:Hpdf}
		f_{\abs{\mathcal{H}}}\left(\abs{h}\right)=\dfrac{4\left(r_1r_2\right)^{v/2}}{\Gamma\left(m_1\right)\Gamma\left(m_2\right)}|h|^{v-1}K_n\left(2\sqrt{r_1r_2} |h|\right),
	\end{equation}
	where $\Gamma\of{\cdot}$ denotes the Gamma function\cite[eq: 6.1.1]{1964handbook}, $ K_n\left(\cdot\right) $ denotes the modified Bessel function of the second kind with order $ n $\cite[eq: 9.6.1]{1964handbook}, $ r_1=m_1/\Omega_1 $, $ r_2=m_2/\Omega_2$, $ v=m_1+m_2 $, and $ n=|m_1-m_2| $.
	
	In the following, we analyze two cases with different power allocation schemes, i.e., fixed transmit power and varying transmit power with an average power constraint.
	\subsubsection{Fixed Transmit Power}
    Considering the fixed symbol energy, the average \ac{SER} in fading channels is calculated as
	\begin{gather}
		\label{eq:pe1}
		\overline{P^D_{\textrm{e}}}=\int_{0}^{\infty}\overline{P^D_{\textrm{e}|h}}\, f_{\abs{\mathcal{H}}}\left(\abs{h}\right) \mathrm{d}|h|.
	\end{gather}
	Substituting (\ref{eq:avgSERAWGN}) and (\ref{eq:Hpdf}) into (\ref{eq:pe1}), and substituting $\abs{h}$ with ${\tilde{h}}/{2\sqrt{r_1r_2}}$, we obtain
	\begin{gather}
	\begin{aligned}
		\label{eq:pe2}
		\overline{P^D_{\textrm{e}}}&=\dfrac{2^{2-v}\sqrt{1/\pi}}{M\Gamma\left(m_1\right)\Gamma\left(m_2\right)}\sum_{a=0}^{M-1}\sum_{t=1}^{N_\textrm{GH}}\omega_{t}I_1,
	\end{aligned}
\end{gather}
where 
\begin{gather}
    \begin{aligned}
    \label{eq:I11}
    I_1={\int_{0}^{\infty}{\left[1-F_{\hat{\mathcal{L}}^D_{\left(a\right)}|\tilde{h}}\left(\sqrt{2}\sigma_a x_{t}+\mu_a\right)\right]} \tilde{h}^{v-1}K_n\left(\tilde{h}\right)\mathrm{d}\tilde{h}}.
    \end{aligned}
\end{gather}

In the following, we compute $I_1$ depending on the value of $n$ since for certain cases the Bessel function in (\ref{eq:I11}) can be expressed in closed-form while for other cases it is challenging to find a closed-form expression.

		\paragraph{$n$ is half-integer}
		We consider the order of the Bessel function as a half-integer, i.e., $n=u+\frac{1}{2}$ for all $u\in \offf{0,1,2,\cdots}$. Thus, the modified Bessel function of the second kind with order $n$ can be expressed in closed form\cite[eq: 8.468]{tableofintegral}
		\begin{equation}
			\label{eq:besselhalfint}
			K_{u+\frac{1}{2}}\left(z\right)=\sqrt{\dfrac{\pi}{2z}}\sum_{k=0}^{u}\dfrac{\left(u+k\right)!}{k!\left(u-k\right)!\left(2z\right)^k}\, e^{-z}.
		\end{equation}
		Substituting (\ref{eq:besselhalfint}) into $I_1$ yields
		\begin{align}
			I_1&=\int_{0}^{\infty}\sqrt{\frac{\pi}{2\tilde{h}}}\, {\left[1-F_{\hat{\mathcal{L}}^D_{\left(a\right)}|\tilde{h}}\left(\sqrt{2}\sigma_a x_{t}+\mu_a\right)\right]} \nonumber\label{eq:IqafterGL1}\\
		&\quad\quad\quad\quad\quad\times\sum_{k=0}^{\left\lfloor n\right\rfloor}\dfrac{\left(\left\lfloor n\right\rfloor+k\right)!}{k!\left(\left\lfloor n\right\rfloor-k\right)!\left(2\tilde{h}\right)^k}		 \tilde{h}^{v-1} e^{-\tilde{h}} \mathrm{d}\tilde{h}.
		\end{align}
		where an a power and an exponential terms $\tilde{h}^{v-1} e^{-\tilde{h}}$ are included in the integrand. Thus, unlike using Gauss-Hermite quadrature in (\ref{eq:GHquad}) that integrates on an integrand with an exponential term $e^{-\hat{l}^2}$, we consider Gauss-Laguerre quadrature to evaluate the integral numerically
		\begin{align}
		    	I_1\overset{}{\approx}\sqrt{\frac{\pi}{2}}\, \sum_{e=1}^{N_\textrm{GL}}\omega_e\dfrac{G_{t}\left(x_e\right)}{\sqrt{x_e}}\sum_{k=0}^{\left\lfloor n\right\rfloor}\dfrac{\left(\left\lfloor n\right\rfloor+k\right)!}{k!\left(\left\lfloor n\right\rfloor-k\right)!\left(2x_e\right)^k}, 		\label{eq:IqafterGL}
		\end{align}
		where $G_{t}\left(x_e\right)$ is given by
		\begin{gather}
			G_{t}\left(x_e\right)=1-\prod_{i=0,i\neq a}^{M-1}\left[1-Q_1\left(\dfrac{\left|x_e\sqrt{E_\textrm{s}}\xi^D_{\left(a,i\right)}\right|}{2\sqrt{r_1r_2}\sigma}, \dfrac{\sqrt{2}\sigma_a x_{t}+\mu_a}{\sigma}\right)\right], \nonumber
		\end{gather}
		$ N_\textrm{GL} $ is the number of function samples, $ x_{e}$ are the roots of the generalized Laguerre polynomial $L^{\left(v-1\right)}_{N_\textrm{GL}}\left(x\right)$, and $\omega_{e} $ are the corresponding weights\cite{rabinowitz1959tables}.

		Substituting (\ref{eq:IqafterGL}) into (\ref{eq:pe2}), we obtain the average \ac{SER} in double Nakagami-m fading channel for the half-integer order $n$
			\begin{align}
			\overline{P^D_{\textrm{e}}}=\dfrac{2^{2/3-v}}{M\Gamma\left(m_1\right)\Gamma\left(m_2\right)}\sum_{a=0}^{M-1}\sum_{t=1}^{N_\textrm{GH}}\sum_{e=1}^{N_\textrm{GL}}\omega_{t}\omega_{e}\dfrac{G_{t}\left(x_{e}\right)}{\sqrt{x_{e}}}\nonumber\\
				\times\sum_{k=0}^{\left\lfloor n\right\rfloor}\dfrac{\left(\left\lfloor 	n\right\rfloor+k\right)!}{k!\left(\left\lfloor n\right\rfloor-k\right)!\left(2x_e\right)^k}.
			\end{align}
		
		\paragraph{$n$ is not half-integer}
		If $n$ is not half-integer, it is challenging to obtain a closed-form expression for the modified Bessel function of the second kind with an order $n$. Thus, we provide an approximation for the Bessel function using the two Bessel functions with adjacent half-integer orders. The approximation is tight according to numerical results. We also assume that $n$, namely, the difference between the shape parameters of two channels, is larger than $1/2$. 
		This is because the tag is typically placed closer to the \ac{Tx} to send signals to a \ac{Rx} farther away. The link with a shorter distance, i.e., the \ac{Tx}-tag link, is more likely to have a \ac{LOS} path, while the tag-\ac{Rx} link is more likely to experience a more severe multipath fading effect. 
		
		For all $n>{1}/{2}$, there exists only one integer $u$ satisfying the condition
		\begin{equation}
			u-\frac{1}{2}<n< u+\frac{1}{2}.
		\end{equation}
		Both $K_{u-1/2}\left(x\right)$ and $K_{u+1/2}\left(x\right)$ can be written in form of elementary functions with an exponential term, as shown in (\ref{eq:besselhalfint}).
		Let us define a closed-form function $K_{u,n}\of{x}$ as
		\begin{gather}
		    \begin{aligned}
		  K_{u,n}\of{x}=\off{\dfrac{K_{u-1/2}\left(x\right)}{K_{u+1/2}\left(x\right)}}^{u-n}
	\sqrt{K_{u-1/2}\left(x\right)K_{u+1/2}\left(x\right)}, 
		    \end{aligned}
		\end{gather}
		The approximation is provided in \cite{besselkbound2017} as
		\begin{align}
		 K_n\of{x}&\approx C_{u,n}\,K_{u,n}\of{x},\label{eq:bound2}
		\end{align}
		where $C_{u,n}$ is a constant given by
		\begin{equation}
    	C_{u,n}=\dfrac{\left(u-{1}/{2}\right)^{u-n+{1}/{2}}\Gamma\left(n\right)}{\Gamma\left(u+{1}/{2}\right)}< 1.
		\end{equation}


		
		Let us also define the summation term in $K_{u-1/2}\left(x\right)$ and $K_{u+1/2}\left(x\right)$ as
		\begin{align}
	\Psi_{u-1/2}\left(x\right)&=\sum_{k=0}^{u-1}\dfrac{\left(u-1+k\right)!}{k!\left(u-1-k\right)!\left(2x\right)^k}, \label{eq:psi-}\\
		\Psi_{u+1/2}\left(x\right)&=\sum_{k=0}^{u}\dfrac{\left(u+k\right)!}{k!\left(u-k\right)!\left(2x\right)^k},\label{eq:psi+}
		\end{align}
		and $\Psi_{u,n}\of{x}$ as
		\begin{gather}\label{eq:psiun}
		    \begin{aligned}
		  \Psi_{u,n}\of{x}=\off{\dfrac{\Psi_{u-1/2}\left(x\right)}{\Psi_{u+1/2}\left(x\right)}}^{u-n}\sqrt{{ \Psi_{u+1/2}\left(x\right)\Psi_{u-1/2}\left(x\right) }}.
		    \end{aligned}
		\end{gather}
		Substituting (\ref{eq:psiun}) into (\ref{eq:bound2}) results in \begin{align}
		 K_n\of{x}&\approx \sqrt{\frac{\pi}{2x}}C_{u,n}\,\Psi_{u,n}\of{x}\,e^{-x},\label{eq:bound4}
		\end{align}
		which has an exponential term so that the generalized Gauss-Laguerre quadrature can be applied to compute $ I_1 $. Following similar steps, the approximation for the average \ac{SER} under double Nakagami-m fading channels is calculated as
		\begin{align}
		\overline{P^D_{\textrm{e}}}&=\dfrac{2^{2/3-v}C_{u,n}}{M\Gamma\left(m_1\right)\Gamma\left(m_2\right)}\sum_{a=0}^{M-1}\sum_{t=1}^{N_\textrm{GH}}\sum_{e=1}^{N_\textrm{GL}}\frac{\omega_{t}{\omega_\textrm{e}} G_{t}\left(x_\textrm{e}\right)}{\sqrt{x_e}/\Psi_{u,n}\of{x_e}},\label{eq:avgSERfadinglow}
		\end{align}
		\subsubsection{Limited Average Power}
		We consider a scenario where
		the transmit power is under certain constraints.
		Also, the symbol energy $E_\textrm{s}\of{\gamma}$ can be adjusted according to the channel condition, or similarly, instantaneous \ac{SNR}. Let us define the instantaneous \ac{SNR} as
		\begin{equation}
		    \gamma=\dfrac{\overline{E_\textrm{s}}|h|^2}{2\sigma^2M}=\tilde{\gamma}\abs{h}^2,
		\end{equation}
		where $\overline{E_\textrm{s}}$ denotes the average symbol energy and $\tilde{\gamma}\triangleq{\overline{E_\textrm{s}}}/{2\sigma^2M}$. The energy constraint is governed by
		\begin{equation}
			\int_{0}^{\infty}E_\textrm{s}\of{\gamma}f_{\Gamma}\of{\gamma}\mathrm{d}\gamma\leq \overline{E_\textrm{s}},
		\end{equation}
		where $f_{\Gamma}\of{\gamma}$ is the \ac{PDF} of the instantaneous \ac{SNR}.\footnote{{Note that the randomness in the \ac{SNR} permits analyzing energy harvesting schemes where the harvested energy can be considered as a \ac{r.v.}.}}
		It is determined only by the channel power gain $\abs{h}^2$. Thus, we have
		\begin{equation}
			p_{\Gamma}\left(\gamma\right)=\dfrac{2\left(r_1r_2/\tilde{\gamma}\right)^{v/2}}{\Gamma\left(m_1\right)\Gamma\left(m_2\right)}\gamma^{v/2-1}K_n\left(2\sqrt{r_1r_2\gamma/\tilde{\gamma}} \right). 
		\end{equation}
		
		We adopt water-filling scheme, the optimal power allocation scheme, to adjust the transmit power \cite{goldsmith2005wireless}. The scheme is governed by
		\begin{gather}
			\label{eq:waterfilling}
			\dfrac{E_\textrm{s}\left(\gamma\right)}{\overline{E_\textrm{s}}}=\left\{
			\begin{aligned}
				\frac{1}{\gamma_0}-\frac{1}{\gamma},\quad\quad&\gamma>\gamma_0\\
				\quad\quad0\quad,\quad\quad&\gamma\leq\gamma_0
			\end{aligned}
			\right.
		\end{gather}
		where the outage \ac{SNR} $\gamma_0$ is found by numerically solving
		\begin{equation}
		    \int_{\gamma_0}^{\infty}\of{\frac{1}{\gamma_0}-\frac{1}{\gamma}}p\of{\gamma}\mathrm{d}\gamma=1.
		\end{equation}
		Substituting (\ref{eq:waterfilling}) into (\ref{eq:receivedsignal}) yields the expression of the received signal with power allocation
		\begin{equation}
			{r_a}\left[k\right]=\sqrt{\dfrac{\overline{E_\textrm{s}}|h|^2}{\gamma_0}-2\sigma^2M}\, {x_a}\left[k\right]+{\omega}\left[k\right].
		\end{equation}
		Following similar derivation, the average \ac{SER} in double fading channels with water-filling scheme is calculated as
			\begin{align}
			\overline{P^D_{\textrm{e}}}=
			\frac{1}{{\int_{h_0}^{\infty}f_{\abs{\mathcal{H}}}\left(\abs{h}\right)\mathrm{d}|h|}}\int_{h_0}^{\infty}\int_{0}^{\infty}\left[1-F_{\hat{\mathcal{L}}^D_{\left(a\right)}|h}\left(l\right)\right] \nonumber\\
			\times f_{\mathcal{L}^D_{\left(a,a\right)}|h}\left(l\right){f_{\abs{\mathcal{H}}}\left(\abs{h}\right)}\mathrm{d}l\ \mathrm{d}|h|, 
		\end{align}
		where $h_0$ is the amplitude of the outage channel amplitude defined as $h_0=\sqrt{{\gamma_0}/{\tilde{\gamma}}}$. The integral is divided by ${\int_{h_0}^{\infty}f_{\abs{\mathcal{H}}}\left(\abs{h}\right)\mathrm{d}|h|}$ since it is a conditional probability that the \ac{SNR} is higher than the outage.

\section{Numerical Results and Discussion}
\label{sec:numerical}

In this section, we show the performance of \ac{LB} in terms of the power spectrum and error performance in \ac{AWGN} and fading channels.
\subsection{Power Spectrum}
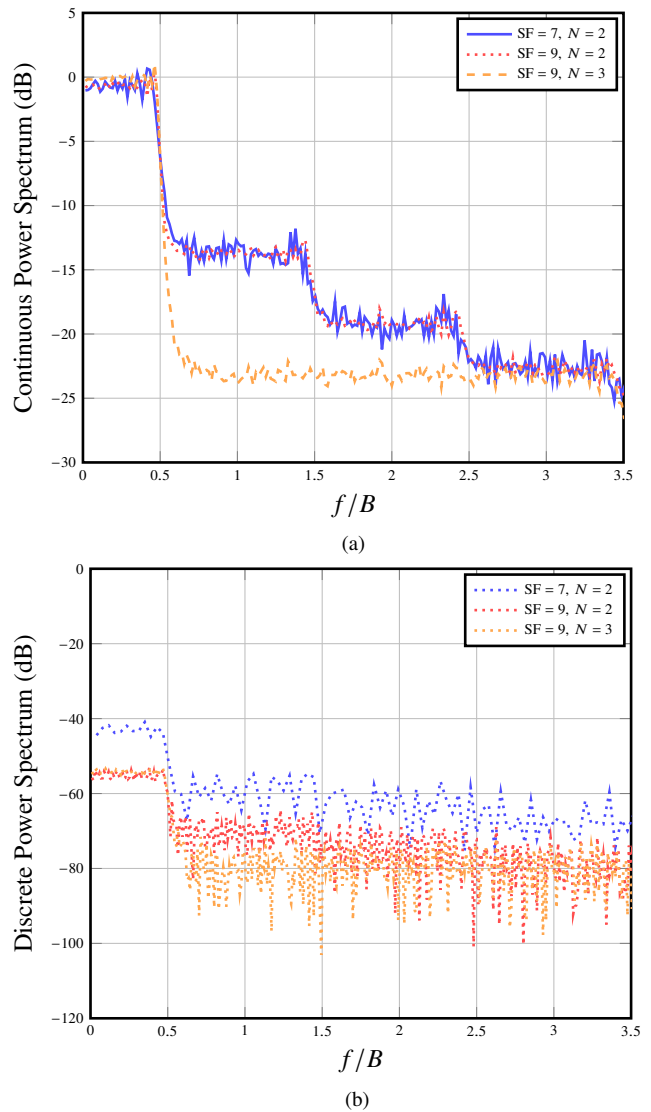
\begin{figure}[t]
	\centering 
	\subfloat[\hspace*{-4.5em}]{
		\label{fig:PSDContinous}
		\pgfplotsset{every axis/.append style={
		font=\footnotesize,
		line width=1pt,
		legend style={font=\footnotesize, at={(0.99,0.99)}},legend cell align=left,nodes={scale=0.7, transform shape}},
} %
\pgfplotsset{compat=1.13}
\begin{tikzpicture}
	\begin{axis}[
		xlabel near ticks,
		ylabel near ticks,
		grid=both,
		xlabel={$f/B$},
		ylabel={Continuous Power Spectrum (dB)},
		ytick={5,0,-5,-10,-15,-20,-25,-30},
		yticklabel style={/pgf/number format},
		width=0.99\linewidth,
		xmin= 0, xmax=3.5,
		ymin=-30, ymax=5,
		ylabel style={font=\Large},
		xlabel style={font=\Large},
		]
		\addplot[blue!70!white]table {Figures/PSD/Continuous/PSDConDataSF7NQ2.dat};
		\addlegendentry{$\text{SF}=7$, $N=2$};
		
		\addplot[red!70!white,dotted]table {Figures/PSD/Continuous/PSDConDataSF9NQ2.dat};
		\addlegendentry{$\text{SF}=9$, $N=2$};
		
		\addplot[orange!70!white,dashed]table {Figures/PSD/Continuous/PSDConDataSF9NQ3.dat};
		\addlegendentry{$\text{SF}=9$, $N=3$};
	\end{axis}
\end{tikzpicture}}
	\qquad
	\subfloat[\hspace*{-4.5em}]{
		\label{fig:PSDDiscrete}
		\pgfplotsset{every axis/.append style={
		font=\footnotesize,
		line width=1pt,
		legend style={font=\footnotesize, at={(0.99,0.99)}},legend cell align=left,nodes={scale=0.7, transform shape}},
} %
\pgfplotsset{compat=1.13}
\begin{tikzpicture}
	\begin{axis}[
		xlabel near ticks,
		ylabel near ticks,
		grid=both,
		xlabel={$f/B$},
		ylabel={Discrete Power Spectrum (dB)},
		ytick={0,-20,-40,-60,-80,-100,-120},
		yticklabel style={/pgf/number format},
		width=0.99\linewidth,
		xmin= 0, xmax=3.5,
		ymin=-120, ymax=0,
		ylabel style={font=\Large},
		xlabel style={font=\Large},
		]
		\addplot[blue!70!white,dotted]table {Figures/PSD/Discrete/PSDDisDataSF7NQ2.dat};
		\addlegendentry{$\text{SF}=7$, $N=2$};
		
		\addplot[red!70!white,dotted]table {Figures/PSD/Discrete/PSDDisDataSF9NQ2.dat};
		\addlegendentry{$\text{SF}=9$, $N=2$};
		
		\addplot[orange!70!white,dotted]table {Figures/PSD/Discrete/PSDDisDataSF9NQ3.dat};
		\addlegendentry{$\text{SF}=9$, $N=3$};
	\end{axis}
\end{tikzpicture}}
	\caption{Continuous and discrete spectrum of baseband \acl{LB} signals with $\text{SF}\in\offf{7,9}$ and $N\in\offf{2,3}$: (a) continuous part of the spectrum, (b) discrete spectrum.}
	\label{fig:PSDCONvsDIS}
\end{figure}

In Fig.~\ref{fig:PSDCONvsDIS}, we present the double-sided power spectrum of the baseband \ac{LB} signals as a function of normalized frequency $f/B$ with different spreading factors and the number of phases, i.e., $\text{SF}\in\offf{7,9}$ and $N\in\offf{2,3}$. We only show $\mathcal{G}_{\mathcal{I}}\of{f}$ for $f\geq0$ because $\mathcal{G}_{\mathcal{I}}\of{f}=\mathcal{G}_{\mathcal{I}}\of{-f}$.
We show both the normalized power spectral density,  $10\log\of{\mathcal{G}_\mathcal{I}^{\of{c}}\of{f}*B}$, and the discrete part of the spectrum. For the discrete spectrum, we show the power $\abs{\sum_{a=0}^{M-1}S_a\of{\frac{lB}{M}}}^2/{M^2T^2_\text{s}}$ at frequency $lB/M$. We see from the continuous part of the spectrum that it has a staircase shape where the power spectrum does not always decrease as frequency increases but maintains relatively stable over some frequency ranges. 
Also,  for a higher number of phases, i.e., loads,  the spectrum drops faster compared to that with a smaller number of quantization phases.  
We can also see that as SF increases, the power spectrum gets increasingly condensed so that more power of the complex envelope is contained between $-B/2$ and $B/2$.



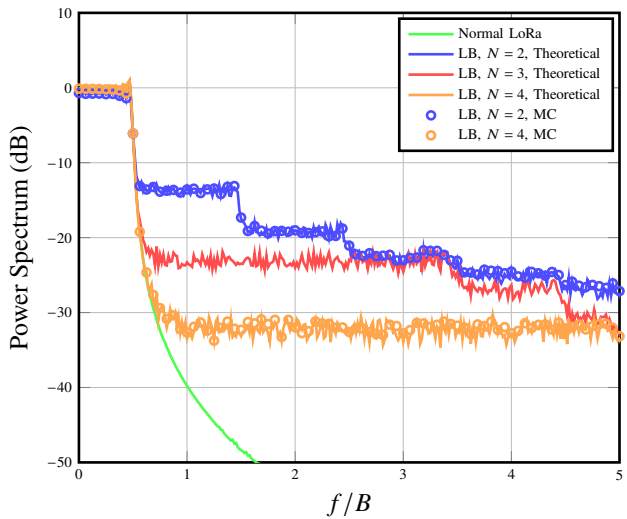
\begin{figure}[t]
	\centering
	\pgfplotsset{every axis/.append style={
		font=\footnotesize,
		line width=1pt,
		legend style={font=\footnotesize, at={(0.99,0.99)}},legend cell align=left,nodes={scale=0.7, transform shape}},
} %
\pgfplotsset{compat=1.13}
	\begin{tikzpicture}
\begin{axis}[
xlabel near ticks,
ylabel near ticks,
grid=both,
xlabel={$f/B$},
ylabel={Power Spectrum (dB)},
	ytick={10,0,-10,-20,-30,-40,-50,-60,-70},
yticklabel style={/pgf/number format},
width=0.99\linewidth,
	xmin= 0, xmax=5,
	ymin=-50, ymax=10,
ylabel style={font=\Large},
xlabel style={font=\Large},
]

\addplot[green!70!white]table {Figures/PSD/PowerSpectrum/PSMonteCarloDataSF9LoRa.dat};
\addlegendentry{Normal LoRa};

\addplot[blue!70!white]table {Figures/PSD/PowerSpectrum/PSDataSF9NQ2.dat};
\addlegendentry{\ac{LB}, $N=2$, Theoretical};

\addplot[red!70!white]table {Figures/PSD/PowerSpectrum/PSDataSF9NQ3.dat};
\addlegendentry{\ac{LB}, $N=3$, Theoretical};

\addplot[orange!70!white]table {Figures/PSD/PowerSpectrum/PSDataSF9NQ4.dat};
\addlegendentry{\ac{LB}, $N=4$, Theoretical};


\addplot[blue!70!white,only marks,mark=o,mark size=1.5, mark repeat=4]table {Figures/PSD/PowerSpectrum/PSMonteCarloDataSF9NQ2.dat};
\addlegendentry{\ac{LB}, $N=2$, MC};


\addplot[orange!70!white,only marks,mark=o,mark size=1.5,mark repeat=4]table {Figures/PSD/PowerSpectrum/PSMonteCarloDataSF9NQ4.dat};
\addlegendentry{\ac{LB}, $N=4$, MC};

\end{axis}
\end{tikzpicture}
	\caption{{Power Spectrum of baseband LoRa and \acl{LB} modulated signals at $\text{SF}=9$ for $N\in\offf{2,3,4}$.}}
	\label{fig:PowerSpectrum}
\end{figure}

In Fig.~\ref{fig:PowerSpectrum}, we compare the derived total power spectrum with both continuous and discrete parts with that of normal LoRa at $\text{SF}=9$, for various numbers of quantization phases, i.e., $N\in\offf{2,3,4}$. The frequency range of interest is split into multiple bins of $\Delta f = B/1024$ width. 
{We verified the analytical results using Monte Carlo simulations generated by estimating the power spectral density in \eqref{eq:randomprocess} using Welch's method. Only the cases for $N\in\offf{2,4}$ are included for space limitation.} 
We see from the figure that the power spectrum of \ac{LB} modulated signals has a strong similarity with that of normal LoRa in the range of $f\in\off{0,B/2}$.
However, the power spectrum of \ac{LB} signals begins to saturate for $f>B/2$ while that of normal LoRa decreases continuously. For an increasing number of phases, \ac{LB} matches normal LoRa for a larger frequency range. 
For example, at $f/B=1$, the power spectrum of \ac{LB} with $N=5$ is nearly equivalent to normal LoRa, while for $N=2$, the gap is approximately over 25~dB.

\begin{figure}[t]
	\centering
	\input{Figures/PSD/Mask/Figure_PSMASK}
	\caption{One-sided power spectrum for \acl{LB} passband signals compared with the mask of the ETSI regulation in the G1 sub-band, for $\text{SF}=7$, $N\in\offf{3,4}$, $B= 250~\text{kHz}$, $\text{RBW=1~kHz}$ and $P_s=14 ~\text{dBm}$.}
	\label{fig:PowerSpectrumMask}
\end{figure}

Also, we investigate the \ac{LB} spectrum in the ISM band with the ETSI regulations\cite{ETSI}. In Fig.~\ref{fig:PowerSpectrumMask}, we report the one-sided power spectrum calculated with $\text{SF}=7$, the resolution bandwidth $\text{RBW}=1~\text{kHz}$ and transmission power $P_s=\text{14~dBm}$, i.e., the maximum allowed power. The spectrum exceeds the spectral mask for $N=3$ while it is within the mask for $N=4$ at the maximum transmission power. Thus, to satisfy the ETSI regulation on the ISM band, one should either increase the number of loads used in the tag or decrease the transmission power. The same approach may be used to investigate the compliance of \ac{LB} for different ISM bands, spreading factors, number of quantization phases, and bandwidths in accordance with other regional requirements.

\subsection{Error Performance in \ac{AWGN} Channels}
\begin{figure}[t]
	\centering
	\pgfplotsset{every axis/.append style={
		font=\footnotesize,
		line width=1pt,
		legend style={font=\footnotesize, at={(0.295,0.58)}},legend cell align=left,nodes={scale=0.7, transform shape}},
} %
\pgfplotsset{compat=1.13}
	\begin{tikzpicture}
\begin{axis}[
ymode=log,
xlabel near ticks,
ylabel near ticks,
grid=both,
xlabel={SNR},
ylabel={SER},
yticklabel style={/pgf/number format},
width=0.99\linewidth,
legend entries={
    $\mathrm{SF}=7$ LoRa,
	$\mathrm{SF}=7$ LB-2,
	$\mathrm{SF}=7$ LB-3,
	$\mathrm{SF}=7$ LB-4,
	$\mathrm{SF}=8$ LoRa,
	$\mathrm{SF}=8$ LB-2,
	$\mathrm{SF}=8$ LB-3,
	$\mathrm{SF}=8$ LB-4,
	$\mathrm{SF}=9$ LoRa,
	$\mathrm{SF}=9$ LB-2,
	$\mathrm{SF}=9$ LB-3,
	$\mathrm{SF}=9$ LB-4
	},
	xmin= -18, xmax=-7,
	ymin=1e-4, ymax=1e0,
ylabel style={font=\Large},
xlabel style={font=\Large},
]
\addplot[blue!60!white,solid] table {Figures/LoRa_LB_ML/SERSF7NumLoRa.dat};
\addplot[blue!60!white,only marks,mark=o,mark size=2,mark repeat=2] table {Figures/LoRa_LB_ML/SERSF7MCML2.dat};
\addplot[blue!60!white,only marks,mark=triangle,mark size=2,mark repeat=2] table {Figures/LoRa_LB_ML/SERSF7MCML3.dat};
\addplot[blue!60!white,only marks,mark=asterisk,mark size=2,mark repeat=2] table {Figures/LoRa_LB_ML/SERSF7MCML4.dat};

\addplot[green!60!black,solid] table {Figures/LoRa_LB_ML/SERSF8NumLoRa.dat};
\addplot[green!60!black,only marks,mark=o,mark size=2,mark repeat=2] table {Figures/LoRa_LB_ML/SERSF8MCML2.dat};
\addplot[green!60!black,only marks,mark=triangle,mark size=2,mark repeat=2] table {Figures/LoRa_LB_ML/SERSF8MCML3.dat};
\addplot[green!60!black,only marks,mark=asterisk,mark size=2,mark repeat=2] table {Figures/LoRa_LB_ML/SERSF8MCML4.dat};

\addplot[red!60!white,solid] table {Figures/LoRa_LB_ML/SERSF9NumLoRa.dat};
\addplot[red!60!white,only marks, mark=o,mark size=2, mark repeat=2] table {Figures/LoRa_LB_ML/SERSF9MCML2.dat};
\addplot[red!60!white,only marks,mark=triangle,mark size=2,mark repeat=2] table {Figures/LoRa_LB_ML/SERSF9MCML3.dat};
\addplot[red!60!white,only marks,mark=asterisk,mark size=2,mark repeat=2] table {Figures/LoRa_LB_ML/SERSF9MCML4.dat};




\end{axis}
\end{tikzpicture}
	\caption{{The \ac{SER} performance in the  \ac{AWGN} channel using \ac{ML} decoder at $\text{SF}\in\offf{7,8,9}$ of normal LoRa and \ac{LB} for $N\in\offf{2,3,4}$.} }
	\label{fig:lora_comparison}
\end{figure}
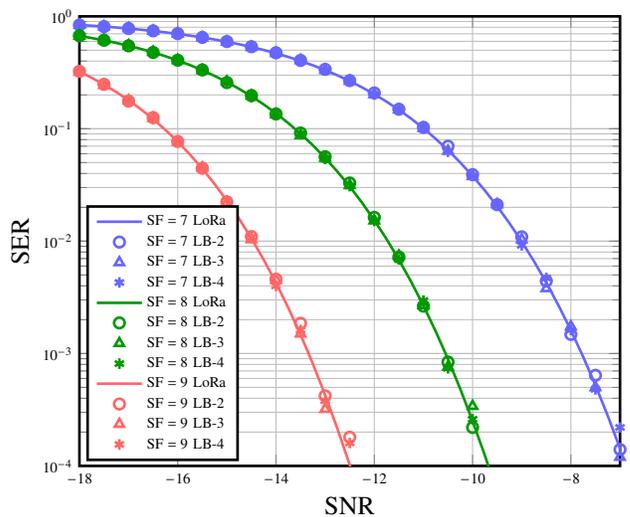

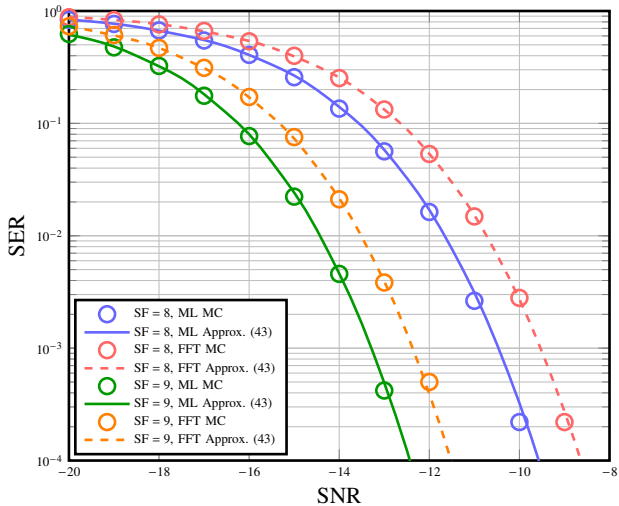
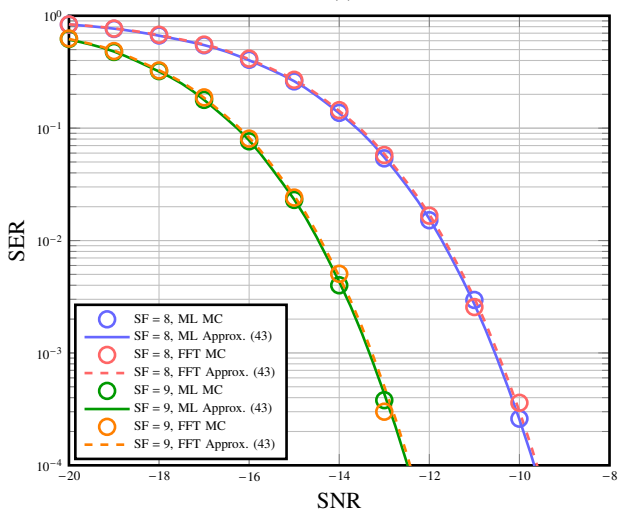
\begin{figure}[t]
	\centering 
	\subfloat[\hspace*{-4.5em}]{
			\label{Fig.awgn2}
			\pgfplotsset{every axis/.append style={
		font=\footnotesize,
		line width=1pt,
		legend style={font=\footnotesize, at={(0.405,0.36)}},legend cell align=left,nodes={scale=0.62, transform shape}},
} %
\pgfplotsset{compat=1.13}
	\begin{tikzpicture}
\begin{axis}[
ymode=log,
xlabel near ticks,
ylabel near ticks,
grid=both,
xlabel={SNR},
ylabel={SER},
width=0.99\linewidth,
yticklabel style={/pgf/number format},
width=0.99\linewidth,
	xmin= -20, xmax=-8,
	ymin=1e-4, ymax=1e0,
ylabel style={font=\Large},
xlabel style={font=\Large},
]

\addplot[blue!60!white,only marks,mark=o,mark size=3,mark repeat=4] table {Figures/SER_AWGN/SERSF8MCML2.dat};
\addlegendentry{$\mathrm{SF}=8$, ML MC}

\addplot[blue!60!white,solid] table {Figures/SER_AWGN/SERSF8AppML2.dat};
\addlegendentry{$\mathrm{SF}=8$, ML Approx. (\ref{eq:avgSERAWGN})}


\addplot[red!60!white,only marks,mark=o,mark size=3,mark repeat=4] table {Figures/SER_AWGN/SERSF8MCfft2.dat};
\addlegendentry{$\mathrm{SF}=8$, FFT MC}

\addplot[red!60!white,dashed] table {Figures/SER_AWGN/SERSF8Appfft2.dat};
\addlegendentry{$\mathrm{SF}=8$, FFT Approx. (\ref{eq:avgSERAWGN})}


\addplot[green!60!black,only marks,mark=o,mark size=3,mark repeat=4] table {Figures/SER_AWGN/SERSF9MCML2.dat};
\addlegendentry{$\mathrm{SF}=9$, ML MC}

\addplot[green!60!black,solid] table {Figures/SER_AWGN/SERSF9AppML2.dat};
\addlegendentry{$\mathrm{SF}=9$, ML Approx. (\ref{eq:avgSERAWGN})}


\addplot[orange,only marks,mark=o,mark size=3,mark repeat=4] table {Figures/SER_AWGN/SERSF9MCfft2.dat};
\addlegendentry{$\mathrm{SF}=9$, FFT MC}

\addplot[orange,dashed] table {Figures/SER_AWGN/SERSF9Appfft2.dat};
\addlegendentry{$\mathrm{SF}=9$, FFT Approx. (\ref{eq:avgSERAWGN})}




\end{axis}
\end{tikzpicture}}
	\qquad
	\subfloat[\hspace*{-4.5em}]{
			\label{Fig.awgn4}
			\pgfplotsset{every axis/.append style={
		font=\footnotesize,
		line width=1pt,
		legend style={font=\footnotesize, at={(0.405,0.36)}},legend cell align=left,nodes={scale=0.62, transform shape}},
} %
\pgfplotsset{compat=1.13}
	\begin{tikzpicture}
\begin{axis}[
ymode=log,
xlabel near ticks,
ylabel near ticks,
grid=both,
xlabel={SNR},
ylabel={SER},
width=0.99\linewidth,
yticklabel style={/pgf/number format},
width=0.99\linewidth,
	xmin= -20, xmax=-8,
	ymin=1e-4, ymax=1e0,
ylabel style={font=\Large},
xlabel style={font=\Large},
]
\addplot[blue!60!white,only marks,mark=o,mark size=3,mark repeat=4] table {Figures/SER_AWGN/SERSF8MCML4.dat};
\addlegendentry{$\mathrm{SF}=8$, ML MC}

\addplot[blue!60!white,solid] table {Figures/SER_AWGN/SERSF8AppML4.dat};
\addlegendentry{$\mathrm{SF}=8$, ML Approx. (\ref{eq:avgSERAWGN})}


\addplot[red!60!white,only marks,mark=o,mark size=3,mark repeat=4] table {Figures/SER_AWGN/SERSF8MCfft4.dat};
\addlegendentry{$\mathrm{SF}=8$, FFT MC}

\addplot[red!60!white,dashed] table {Figures/SER_AWGN/SERSF8Appfft4.dat};
\addlegendentry{$\mathrm{SF}=8$, FFT Approx. (\ref{eq:avgSERAWGN})}


\addplot[green!60!black,only marks,mark=o,mark size=3,mark repeat=4] table {Figures/SER_AWGN/SERSF9MCML4.dat};
\addlegendentry{$\mathrm{SF}=9$, ML MC}

\addplot[green!60!black,solid] table {Figures/SER_AWGN/SERSF9AppML4.dat};
\addlegendentry{$\mathrm{SF}=9$, ML Approx. (\ref{eq:avgSERAWGN})}


\addplot[orange,only marks,mark=o,mark size=3,mark repeat=4] table {Figures/SER_AWGN/SERSF9MCfft4.dat};
\addlegendentry{$\mathrm{SF}=9$, FFT MC}

\addplot[orange,dashed] table {Figures/SER_AWGN/SERSF9Appfft4.dat};
\addlegendentry{$\mathrm{SF}=9$, FFT Approx. (\ref{eq:avgSERAWGN})}




\end{axis}
\end{tikzpicture}}
	\caption{{\ac{LB} \ac{SER} performance in the \ac{AWGN} channel at \text{SF}=8,9: (a) $N=2$, (b) $N=4$.}}
		\label{fig:SERAWGN}
\end{figure}
%

In Fig.~\ref{fig:lora_comparison}, we compare the \ac{SER} using \ac{ML} decoder in the \ac{AWGN} channel at $\text{SF}\in\offf{7,8,9}$ for normal LoRa and \ac{LB} with the different number of quantization phases, where, as a benchmark, the curves for normal LoRa are calculated by computing integrals numerically:
\begin{equation}
	\label{eq:LoRaSER}
	\int_0^\infty\left[1-\left[1-\exp\left[-\frac{l^2}{2\sigma^2}\right]^{M-1}\right]\right]\, f_{\mathcal{L}_{\of{a,a}}^{\text{ML}}}\left(l\right)\mathrm{d}l,
\end{equation}
and the curves for \ac{LB} are from Monte Carlo simulation. 
{It can be observed that the \ac{SER} for \ac{LB} using \ac{ML} decoder has a strong similarity to normal LoRa for various values of the SFs and  number of loads, even though  the \ac{LB} waveforms are non-orthogonal. We verified the results for small lower values of $\mathrm{SF}$ and still find the error performance similar to each other. The results suggest that existing works discussing the error performance of orthogonal LoRa can be also applied to \ac{LB} using \ac{ML} when considering detectors.}

Fig.~\ref{fig:SERAWGN} presents comparisons of derived \ac{LB} \ac{SER} approximation in (\ref{eq:avgSERAWGN}) with the theoretical \ac{SER} calculated by numerically computing the integral in (\ref{eq:peaint}) at $\text{SF}=8,9$ in the \ac{AWGN} channel. The figures also include the results from the Monte Carlo simulation. As shown in the figure, the derived expression in (\ref{eq:avgSERAWGN}) exhibits a tight approximation to the theoretical results. Furthermore, comparing the \ac{SER} using \ac{ML} and \ac{FFT} decoders, the \ac{SER} gap between them differs hugely depending on the number of quantization phases. Improved \ac{SER} using \ac{FFT} decoder is evident when increasing $N$. 
{While the gap is around $1$ dB at \ac{SER} of $10^{-3}$ for $\text{SF}=9$ when using only 4 quantization phases, there are nearly no differences between the performance of two decoders with $16$ quantization phases, indicating that the error performance with $N\geq4$ is similar to orthogonal LoRa for both \ac{ML} and \ac{FFT} detectors in \ac{AWGN} channel.}

\subsection{Error Performance in Fading Channels With Fixed Transmit Power and Limited Average Power}
\begin{figure}[t]
	\centering
	\pgfplotsset{every axis/.append style={
		font=\footnotesize,
		line width=1pt,
		legend style={font=\footnotesize, at={(0.34,0.5)}},legend cell align=left,nodes={scale=0.6, transform shape}},
} %
\pgfplotsset{compat=1.13}
	\begin{tikzpicture}
\begin{axis}[
ymode=log,
xlabel near ticks,
ylabel near ticks,
grid=both,
xlabel={SNR},
ylabel={SER},
yticklabel style={/pgf/number format}, 
width=0.99\linewidth,
	xmin= 5, xmax=25,
	ymin=1e-4, ymax=1e0,
ylabel style={font=\Large},
xlabel style={font=\Large},
]
\node[anchor=west] (source) at (axis cs:18,7e-1){Tag moving closer to Tx};
\node (destination) at (axis cs:15,1e-3){};
\draw[->](source)--(destination);


\addplot[blue!60!white,only marks,mark=o,mark size=3,mark repeat=4] table {Figures/fading_distance/SERMCfftr12.dat};
\addlegendentry{$d_2=\,\quad d_1$, MC}

\addplot[blue!60!white,dashed] table {Figures/fading_distance/SERAPPlowfftr12.dat};
\addlegendentry{$d_2=\,\quad d_1$, Approx. }



\addplot[blue!60!black,only marks,mark=o,mark size=3,mark repeat=4] table {Figures/fading_distance/SERMCfftr14.dat};
\addlegendentry{$d_2=\,\quad d_1$, MC}

\addplot[blue!60!black,solid] table {Figures/fading_distance/SERAPPlowfftr14.dat};
\addlegendentry{$d_2=\,\quad d_1$, Approx. }


\addplot[green!50!white,only marks,mark=o,mark size=3,mark repeat=4] table {Figures/fading_distance/SERMCfftr42.dat};
\addlegendentry{$d_2=\,\,4\,\,d_1$, MC}
\addplot[green!50!white,dashed] table {Figures/fading_distance/SERAPPlowfftr42.dat};
\addlegendentry{$d_2=\,\,4\,\,d_1$, Approx. }


\addplot[green!60!black,only marks,mark=o,mark size=3,mark repeat=4] table {Figures/fading_distance/SERMCfftr44.dat};
\addlegendentry{$d_2=\,\,4\,\,d_1$, MC}
\addplot[green!60!black,solid] table {Figures/fading_distance/SERAPPlowfftr44.dat};
\addlegendentry{$d_2=\,\,4\,\,d_1$, Approx. }

\addplot[red!60!white,only marks,mark=o,mark size=3,mark repeat=4] table {Figures/fading_distance/SERMCfftr162.dat};
\addlegendentry{$d_2=16\,d_1$, MC}
\addplot[red!60!white,dashed] table {Figures/fading_distance/SERAPPlowfftr162.dat};
\addlegendentry{$d_2=16\,d_1$, Approx. }


\addplot[red!60!black,only marks,mark=o,mark size=3,mark repeat=4] table {Figures/fading_distance/SERMCfftr164.dat};
\addlegendentry{$d_2=16\,d_1$, MC}
\addplot[red!60!black,solid] table {Figures/fading_distance/SERAPPlowfftr164.dat};
\addlegendentry{$d_2=16\,d_1$, Approx. }




\end{axis}
\end{tikzpicture}
	\caption{{\ac{LB} \ac{SER} performance in double Nakagami-m fading channels using \ac{ML} decoder for $\textrm{SF}=7$ and $N=2$ (dashed), $N=4$ (solid).} }
	\label{fig:fading_fig}
\end{figure}
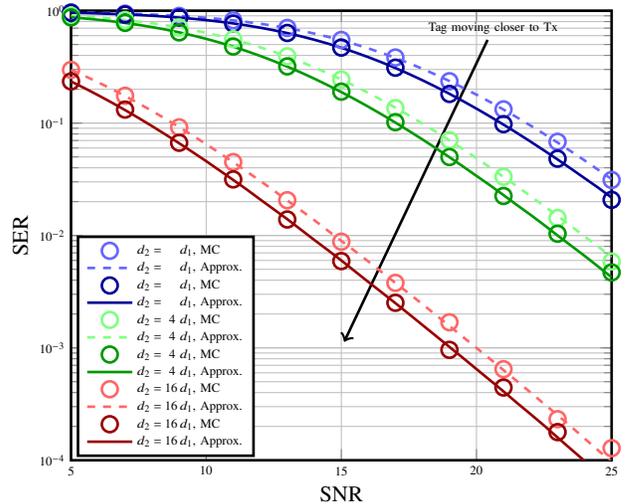
\begin{figure}[t]
	\centering
	\pgfplotsset{every axis/.append style={
		font=\footnotesize,
		line width=1pt,
		legend style={font=\footnotesize, at={(0.415,0.4)}},legend cell align=left,nodes={scale=0.7, transform shape}},
} %
\pgfplotsset{compat=1.13}
	\begin{tikzpicture}
\begin{axis}[
ymode=log,
xlabel near ticks,
ylabel near ticks,
grid=both,
xlabel={SNR},
ylabel={SER},
yticklabel style={/pgf/number format},
width=0.99\linewidth,
	xmin= 5, xmax=25,
	ymin=1e-3, ymax=1e-1,
ylabel style={font=\Large},
xlabel style={font=\Large},
]
%

\addplot[red!60!white,only marks,mark=asterisk,mark size=2,mark repeat=4] table {Figures/SER_water_filling/SERWFMCSF7N2.dat};
\addlegendentry{$\text{SF=7}$, $N=2$, MC}

\addplot[red!60!white,dashed] table {Figures/SER_water_filling/SERWFNumSF7N2.dat};
\addlegendentry{$\text{SF=7}$, $N=2$, Num (\ref{eq:waterfilling})}

\addplot[red!60!white,only marks,mark=o,mark size=2,mark repeat=4] table {Figures/SER_water_filling/SERWFMCSF7N4.dat};
\addlegendentry{$\text{SF=7}$, $N=4$, MC}

\addplot[red!60!white,solid] table {Figures/SER_water_filling/SERWFNumSF7N4.dat};
\addlegendentry{$\text{SF=7}$, $N=4$, Num (\ref{eq:waterfilling})}

\addplot[green!60!black,only marks,mark=asterisk,mark size=2,mark repeat=4] table {Figures/SER_water_filling/SERWFMCSF8N2.dat};
\addlegendentry{$\text{SF=8}$, $N=2$, MC}

\addplot[green!60!black,dashed] table {Figures/SER_water_filling/SERWFNumSF8N2.dat};
\addlegendentry{$\text{SF=8}$, $N=2$, Num (\ref{eq:waterfilling})}

\addplot[green!60!black,only marks,mark=o,mark size=2,mark repeat=4] table {Figures/SER_water_filling/SERWFMCSF8N4.dat};
\addlegendentry{$\text{SF=8}$, $N=4$, MC}

\addplot[green!60!black,solid] table {Figures/SER_water_filling/SERWFNumSF8N4.dat};
\addlegendentry{$\text{SF=8}$, $N=4$, Num (\ref{eq:waterfilling})}

\end{axis}
\end{tikzpicture}
	\caption{{\ac{SER} performance for \ac{LB} with water-filling technique using \ac{FFT} decoder at $\text{SF}\in\offf{7,8}$ and $N\in\offf{2,4}$.}}
	\label{fig:wf}
\end{figure}
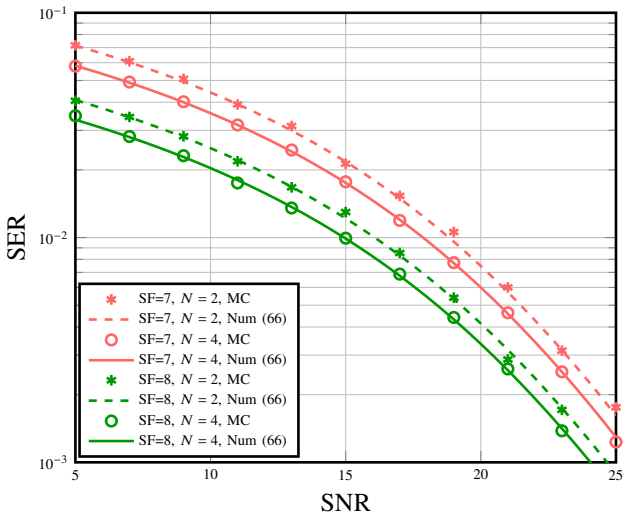
In Fig.~\ref{fig:fading_fig}, we plot the \ac{SER} performance of \ac{LB} over double Nakagami-m fading channels at $\text{SF}=7$ and $N=2${$,4$} where we fix the total distance of the reflected path between the Tx and Rx through the tag, i.e., $d=d_1+d_2$. 
{In addition, we consider a path loss model where the energy is inversely proportional to the squared distance, i.e., $\mathbb{V}\off{\abs{h_1}}\varpropto 1/d_1^2$ and $\mathbb{V}\off{\abs{h_2}}\varpropto 1/d_2^2$.} 
{In this simulation, we consider the \ac{FFT} decoder.} 
It is shown that a better \ac{SER} performance is achieved when the tag is placed closer to the \ac{Tx}, which is similar to the experimental results in \cite{3lorabsk}. Also, the comparison of the derived \ac{SER} approximation (\ref{eq:avgSERfadinglow}) to the curves calculated by numerical integrating (\ref{eq:pe1}) and the Monte Carlo simulation presents an accurate approximation.

On the other hand, for the model with limited average power, we show the error performance of \ac{LB} with a water-filling power allocation scheme in Fig.~\ref{fig:wf} using \ac{FFT} decoder at $\text{SF}\in\offf{7,8}$ and $N\in\offf{2,4}$.
The theoretical curves from numerical integrating (\ref{eq:waterfilling}) are consistent with the Monte Carlo simulations. 
\section{Conclusion}
\label{sec:conclude}
In this paper, we provide the first mathematical expression of \ac{LB} signals with a finite number of loads. Based on the expression, we derived the closed-form expressions for the power spectrum of \ac{LB}, showing the staircase-shaped spectrum. To satisfy the wireless transmission regulations, measures such as increasing the number of phases or decreasing the transmit power may be needed.
An analytical approximation of \ac{SER} for \ac{LB} in both \ac{AWGN} and double Nakagami-m fading channels using two different decoders is derived. {The results suggest that the \ac{SER} performance of \ac{LB} using \ac{ML} decoder with a small number of phases is similar to LoRa modulation, while the \ac{SER} performance using \ac{FFT} decoder is worse.} 
On the other hand, \ac{SER} performance using \ac{FFT} decoder will improve with more quantization phases. In the double fading scenarios, a longer communication range can be achieved by setting the backscatter tag closer to \ac{Tx}.
\section*{Appendix A\\Proof that \acl{ML} decoder is optimum for \acl{LB}}
The \ac{Tx} sends the real part of the passband signal with symbol $a$ to the \ac{Rx}:
\begin{equation}
	R\left(t\right)=\operatorname{Re}\left[r_a\left(t\right)e^{j2\pi f_\textrm{c}t}\right],
\end{equation}
where $r_a\left(t\right)=\left|h\right|\sqrt{E_\textrm{s}}x_a\left(t\right)e^{j\theta_h}+\omega\left(t\right)$ is the corresponding baseband signal, $\theta_h$ is the phase of complex channel gain $h$, and $f_\textrm{c}$ is the carrier frequency. We assume that the channel amplitude $\abs{h}$ and the delay $\tau$ are known to \ac{Rx}. The symbol decision is based on choosing the maximum among a set of posteriori probabilities\cite{digitalcommunicationoverfadingchannell}
	\begin{align}
		\hat{a}&=\mathop{\arg\max}\limits_{0\leq a \leq M-1} P_a \, p\left( R\left(t\right)\big| x_a\left(t\right), \abs{h},\tau\right)\label{eq:ahat}\\
		&=\mathop{\arg\max}\limits_{0\leq a \leq M-1}  \int_{0}^{2\pi} p\left( R\left(t\right)\big| x_a\left(t\right), \left|h\right|, \tau, \theta_h\right)p\left(\theta_h\right)\mathrm{d} \theta_h\nonumber,
	\end{align}
where $P_a$ is the probability of transmitting symbol $a$, $p\left(\theta_h\right)$ is the \ac{PDF} of $\theta_h$. We consider equiprobable transmit symbols and uniformly distributed phases as is typical for Rayleigh, Rician, and Nakagami fading channels. The probability $ p\left( R\left(t\right)\big| x_a\left(t\right), \left|h\right|,\tau, \theta_h\right)$ is a joint Gaussian \ac{PDF}
	\begin{align}
	&p\left( R\left(t\right)\big| x_a\left(t\right), \left|h\right|,\tau, \theta_h\right)=\nonumber\\
&\quad\quad\quad\quad	K\exp\of{\operatorname{Re}\offf{\frac{\abs{h}}{N_0}e^{-j\theta_h}y_a\of{\tau}}-\frac{\abs{h}^2E_a}{N_0}},
	\end{align}
where $K$ is an integral constant and 
\begin{gather}
	\begin{aligned}
		y_a\of{\tau}=\int_{0}^{T_s}r_a\of{t+\tau}x_a^*\of{t}\mathrm{d} t
	\end{aligned}
\end{gather}
is the complex correlation of the received signal and the signal waveform of symbol $a$ and 
\begin{gather}
	\begin{aligned}
		E_a=\frac{1}{2}\int_{0}^{T_s}\abs{x_a\of{t}}^2\mathrm{d}t
	\end{aligned}
\end{gather}
is the energy of the waveform $x_a\of{t}$. Thus, we obtain
	\begin{align}
	&p\left( R\left(t\right)\big| x_a\left(t\right), \abs{h},\tau\right)=\label{eq:ahatcon}\\
&	\exp\of{-\frac{\abs{h}^2E_a}{N_0}}\frac{K}{2\pi}\int_{0}^{2\pi}\exp\of{\operatorname{Re}\offf{\frac{\abs{h}}{N_0}e^{-j\theta_h}y_a\of{\tau}}}\mathrm{d}\theta_h.\nonumber
	\end{align}
Substitute (\ref{eq:ahatcon}) into (\ref{eq:ahat}) and remove the terms that are independent of $a$, we obtain
	\begin{align}
			\hat{a}&=\mathop{\arg\max}\limits_{0\leq a \leq M-1}\int_{0}^{2\pi}\exp\of{\operatorname{Re}\offf{e^{-j\theta_h}y_a\of{\tau}}}\mathrm{d}\theta_h\nonumber\\
			&=\mathop{\arg\max}\limits_{0\leq a \leq M-1}\int_{0}^{2\pi}\exp\of{\abs{y_a\of{\tau}}\cos\off{\theta_h-\arg\of{y_a\of{\tau}}}}\mathrm{d}\theta_h\nonumber\\
			&=\mathop{\arg\max}\limits_{0\leq a \leq M-1}I_0\of{\abs{y_a\of{\tau}}},
	\end{align}
where the term $E_a$ is independent of $a$ because the waveforms of different symbols have the same energy. Since $I_0$ is a monotonically increasing function, the optimum decision is made by choosing the symbol that has the maximum cross-correlation with the received signal.

\section*{Appendix B\\How to obtain $\offf{t_m}_{m=1}^{\psi}$}

$t_m$ are times when the value of $\widetilde{Q}_N\off{f\of{t}}$ make discrete changes.
The set of $t_m$ can be obtained by solving
\begin{equation}
f\of{t}=\frac{i\pi}{2^{N-1}}, \quad i\in\mathbb{Z}, \quad t\in \off{0,M/B}, 
\end{equation}
where the two parts of $f\of{t}$ are quadratic functions so that the roots can be easily obtained. It is feasible to directly solve $f_1\of{t}={i\pi}/{2^{N-1}}$ and $f_2\of{t}={i\pi}/{2^{N-1}}$ individually in their respective regions. However, it is worth noting that we have the equality
\begin{equation}
    f_2\of{t}=f_1\of{t-\frac{M}{B}}.
\end{equation}
We see that $f_2\of{t}$ is a shifted version of $f_1\of{t}$, which can be leveraged to simplify the solution. More precisely, the roots of $f_2\of{t}={i\pi}/{2^{N-1}}$ in the region $\frac{M-a}{B}\leq t\leq\frac{M}{B} $ are corresponding to the roots of $f_1\of{t}={i\pi}/{2^{N-1}}$ in region $\frac{-a}{B}\leq t\leq 0 $. Hence, all the roots can be obtained by calculating the roots of $f_1\of{t}={i\pi}/{2^{N-1}}$ for $t\in\off{-\frac{a}{B},\frac{M-a}{B}}$ using root formula for quadratic equation
	\begin{align}
	z_{i}^{\pm}=\frac{\of{M-2a}\pm\sqrt{\of{M-2a}^2+i M2^{3-N}}}{2B},\nonumber\\
    \lceil-\frac{\of{M-2a}^2}{2^{3-n}M}\rceil\leq i \leq \lfloor \frac{a\of{M-a}}{2^{1-n}M}\rfloor.
	\end{align}
Next, add $M/B$ to $z_i^{\pm}$ if they are negative. Also, $t=M/B$ should be added to the roots set because it overlaps with root $t=0$ before being shifted. Finally, rearrange all roots from smallest to largest, namely, $t_1\leq t_2 \leq\cdots\leq t_\psi$. 

\section*{Appendix C\\Proof that $\mathcal{L}^{D}_{\left(a,i\right)}$ follow Rician distribution with shape parameter $\kappa^D_{\left(a,i\right)}=\frac{\left|h\sqrt{E_\textrm{s}}\xi^{D}_{\left(a,i\right)}\right|^2}{2\sigma^2}$}
The Rician distribution is the probability distribution of the magnitude of a circularly-symmetric bivariate normal random variable. A \ac{r.v.} $\mathcal{L}=\sqrt{\mathcal{X}^2+\mathcal{Y}^2}$ follows the Rician distribution with shape parameter $\kappa={\left|v\right|^2}/{2\sigma^2}$ if $\mathcal{X} \sim \mathcal{N}\left(v\sin{\theta},\sigma^2\right)$ and $\mathcal{Y} \sim \mathcal{N}\left(v\cos{\theta},\sigma^2\right)$, where $\mathcal{X}$ and $\mathcal{Y}$ are independently distributed normal \acp{r.v.} and $\theta=v/\abs{v}$. $\mathcal{L}^{D}_{\left(a,i\right)}$ is the amplitude of a complex \ac{r.v.}
\begin{align}
	\mathcal{L}^{D}_{\left(a,i\right)}&=\abs{
		h\sqrt{E_\textrm{s}}\xi^{D}_{\left(a,i\right)}+\mathcal{W}^{D}_{\left(a,i\right)
	}}\\
	&=\abs{
		\abs{h\sqrt{E_\text{s}}\xi_{\of{a,i}}^D
		}e^{j\beta}e^{-j\beta}+\mathcal{W}^{D}_{\left(a,i\right)}e^{-j\beta}
	},
\end{align}
where $e^{j\beta}$ is the angle of $h\sqrt{E_\textrm{s}}\xi^{D}_{\left(a,i\right)}$, and the second term is the rotated version of complex Gaussian \ac{r.v.}. Thus, we define
\begin{align}
	\mathcal{X}&=\left|h\sqrt{E_\textrm{s}}\xi^{D}_{\left(a,i\right)}\right|+\operatorname{Re}\off{\mathcal{W}^{D}_{\left(a,i\right)}e^{-j\beta}},\\
	\mathcal{Y}&=\operatorname{Im}\off{\mathcal{W}^{D}_{\left(a,i\right)}e^{-j\beta}},
\end{align}
where $\mathcal{L}^{D}_{\left(a,i\right)}=\abs{\mathcal{X}+j\mathcal{Y}}$. Since $\mathcal{X}\sim \mathcal{N}\of{\left|h\sqrt{E_\textrm{s}}\xi^{D}_{\left(a,i\right)}\right|,\sigma^2}$ and $\mathcal{Y}\sim \mathcal{N}\of{0,\sigma^2}$, $\mathcal{L}^{D}_{\left(a,i\right)}$ follows Rician distribution with shape parameter $\kappa^D_{\left(a,i\right)}=\frac{\left|h\sqrt{E_\textrm{s}}\xi^{D}_{\left(a,i\right)}\right|^2}{2\sigma^2}$.
%
\bibliographystyle{IEEEtran}
\bibliography{reference.bib}

\begin{thebibliography}{10}
\providecommand{\url}[1]{#1}
\csname url@samestyle\endcsname
\providecommand{\newblock}{\relax}
\providecommand{\bibinfo}[2]{#2}
\providecommand{\BIBentrySTDinterwordspacing}{\spaceskip=0pt\relax}
\providecommand{\BIBentryALTinterwordstretchfactor}{4}
\providecommand{\BIBentryALTinterwordspacing}{\spaceskip=\fontdimen2\font plus
\BIBentryALTinterwordstretchfactor\fontdimen3\font minus
  \fontdimen4\font\relax}
\providecommand{\BIBforeignlanguage}[2]{{%
\expandafter\ifx\csname l@#1\endcsname\relax
\typeout{** WARNING: IEEEtran.bst: No hyphenation pattern has been}%
\typeout{** loaded for the language `#1'. Using the pattern for}%
\typeout{** the default language instead.}%
\else
\language=\csname l@#1\endcsname
\fi
#2}}
\providecommand{\BIBdecl}{\relax}
\BIBdecl

\bibitem{BocHeathPopovski:14}
F.~{Boccardi}, R.~W. {Heath}, A.~{Lozano}, T.~L. {Marzetta}, and P.~{Popovski},
  ``Five disruptive technology directions for {5G},'' \emph{IEEE Communications
  Magazine}, vol.~52, no.~2, pp. 74--80, Feb. 2014.

\bibitem{ShuppingAminSlim:20}
S.~Dang, O.~Amin, B.~Shihada, and M.-S. Alouini, ``What should {6G} be?''
  \emph{Nature Electronics}, vol.~3, no.~1, pp. 20--29, 2020.

\bibitem{vailshery_2022}
\BIBentryALTinterwordspacing
L.~S. Vailshery, ``Iot connected devices worldwide 2019-2030,'' Jun 2022.
  [Online]. Available:
  \url{https://www.statista.com/statistics/1183457/iot-connected-devices-worldwide/}
\BIBentrySTDinterwordspacing

\bibitem{AEloramodulation}
M.~Chiani and A.~Elzanaty, ``On the {LoRa} modulation for {IoT}: Waveform
  properties and spectral analysis,'' \emph{IEEE Internet of Things Journal},
  vol.~6, no.~5, pp. 8463--8470, Oct. 2019.

\bibitem{RazKulSoo:17}
U.~Raza, P.~Kulkarni, and M.~Sooriyabandara, ``Low power wide area networks: An
  overview,'' \emph{IEEE Communications Surveys \& Tutorials}, vol.~19, no.~2,
  pp. 855--873, Secondquarter 2017.

\bibitem{PiaElzGioChi:17b}
D.~Pianini, A.~Elzanaty, A.~Giorgetti, and M.~Chiani, ``Emerging distributed
  programming paradigm for cyber-physical systems over {LoRAWAN}s,'' in
  \emph{2018 IEEE Globecom Workshops (GC Wkshps)}, Dec. 2018, pp. 1--6.

\bibitem{s23031698}
\BIBentryALTinterwordspacing
S.~Sobhi, A.~Elzanaty, M.~Y. Selim, A.~M. Ghuniem, and M.~F. Abdelkader,
  ``Mobility of {LoRaWAN} gateways for efficient environmental monitoring in
  pristine sites,'' \emph{Sensors}, vol.~23, no.~3, 2023. [Online]. Available:
  \url{https://www.mdpi.com/1424-8220/23/3/1698}
\BIBentrySTDinterwordspacing

\bibitem{RezTelHer:20}
F.~{Rezaei}, C.~{Tellambura}, and S.~{Herath}, ``Large-scale wireless-powered
  networks with backscatter communications—a comprehensive survey,''
  \emph{IEEE Open Journal of the Communications Society}, vol.~1, pp.
  1100--1130, July 2020.

\bibitem{advancesBSc2021}
C.~Song, Y.~Ding, A.~Eid, J.~G.~D. Hester, X.~He, R.~Bahr, A.~Georgiadis,
  G.~Goussetis, and M.~M. Tentzeris, ``Advances in wirelessly powered
  backscatter communications: From antenna/{RF} circuitry design to printed
  flexible electronics,'' \emph{Proceedings of the IEEE}, vol. 110, no.~1, pp.
  171--192, Jan. 2022.

\bibitem{VanHoaKim:18}
N.~{Van Huynh}, D.~T. {Hoang}, X.~{Lu}, D.~{Niyato}, P.~{Wang}, and D.~I.
  {Kim}, ``Ambient backscatter communications: A contemporary survey,''
  \emph{IEEE Communications Surveys \& Tutorials}, vol.~20, no.~4, pp.
  2889--2922, May 2018.

\bibitem{WanGaiTel:16}
G.~Wang, F.~Gao, R.~Fan, and C.~Tellambura, ``Ambient backscatter communication
  systems: Detection and performance analysis,'' \emph{IEEE Transactions on
  Communications}, vol.~64, no.~11, pp. 4836--4846, Nov. 2016.

\bibitem{LiuSmith:13}
V.~Liu, A.~Parks, V.~Talla, S.~Gollakota, D.~Wetherall, and J.~R. Smith,
  ``Ambient backscatter: Wireless communication out of thin air,'' \emph{ACM
  SIGCOMM Computer Communication Review}, vol.~43, no.~4, pp. 39--50, Oct.
  2013.

\bibitem{9455142}
Y.~H. Al-Badarneh, A.~Elzanaty, and M.-S. Alouini, ``On the performance of
  spectrum-sharing backscatter communication systems,'' \emph{IEEE Internet of
  Things Journal}, vol.~9, no.~3, pp. 1951--1961, Feb. 2022.

\bibitem{4pbsk}
V.~Iyer, V.~Talla, B.~Kellogg, S.~Gollakota, and J.~Smith, ``Inter-technology
  backscatter: Towards internet connectivity for implanted devices,'' in
  \emph{Proceedings of the 2016 ACM SIGCOMM Conference}, ser. SIGCOMM
  '16.\hskip 1em plus 0.5em minus 0.4em\relax New York, NY, USA: Association
  for Computing Machinery, Aug. 2016, p. 356–369.

\bibitem{3lorabsk}
V.~Talla, M.~Hessar, B.~Kellogg, A.~Najafi, J.~R. Smith, and S.~Gollakota,
  ``{LoRa} backscatter: Enabling the vision of ubiquitous connectivity,''
  \emph{Proceedings of the ACM on Interactive, Mobile, Wearable and Ubiquitous
  Technologies}, vol.~1, no.~3, pp. 1--24, Sept. 2017.

\bibitem{selfsustainlora2021}
X.~Tang, G.~Xie, and Y.~Cui, ``Self-sustainable long-range backscattering
  communication using {RF} energy harvesting,'' \emph{IEEE Internet of Things
  Journal}, vol.~8, no.~17, pp. 13\,737--13\,749, Sept. 2021.

\bibitem{fullduplexlorabsc}
M.~Katanbaf, A.~Weinand, and V.~Talla, ``Simplifying backscatter deployment:
  Full-duplex {LoRa} backscatter,'' in \emph{18th USENIX Symposium on Networked
  Systems Design and Implementation (NSDI 21)}, 2021, pp. 955--972.

\bibitem{xorlora}
H.~Li, X.~Tong, Q.~Li, and X.~Tian, ``{XORLoRa}: {LoRa} backscatter
  communication with commodity devices,'' in \emph{2020 IEEE 6th International
  Conference on Computer and Communications (ICCC)}, Dec. 2020, pp. 706--711.

\bibitem{plora}
Y.~Peng, L.~Shangguan, Y.~Hu, Y.~Qian, X.~Lin, X.~Chen, D.~Fang, and
  K.~Jamieson, ``Plora: A passive long-range data network from ambient {LoRa}
  transmissions,'' in \emph{Proceedings of the 2018 Conference of the ACM
  Special Interest Group on Data Communication}, 2018, pp. 147--160.

\bibitem{freeback2021}
G.~Huang, P.~Yang, H.~Zhou, Y.~Yan, X.~He, and X.~Li, ``Freeback: Blind and
  distributed rate adaptation in {LoRa}-based backscatter networks,'' in
  \emph{2021 IEEE Wireless Communications and Networking Conference (WCNC)},
  Mar.-Apr. 2021, pp. 1--6.

\bibitem{onoffkeylorabsc}
X.~Guo, L.~Shangguan, Y.~He, J.~Zhang, H.~Jiang, A.~A. Siddiqi, and Y.~Liu,
  ``Efficient ambient {LoRa} backscatter with on-off keying modulation,''
  \emph{IEEE/ACM Transactions on Networking}, pp. 1--14, Nov. 2021.

\bibitem{wearableLoRa2020}
M.~Lazaro, A.~Lazaro, and R.~Villarino, ``Feasibility of backscatter
  communication using {LoRAWAN} signals for deep implanted devices and wearable
  applications,'' \emph{Sensors}, vol.~20, no.~21, p. 6342, Nov. 2020.

\bibitem{7059230}
N.~Fasarakis-Hilliard, P.~N. Alevizos, and A.~Bletsas, ``Coherent detection and
  channel coding for bistatic scatter radio sensor networking,'' \emph{IEEE
  Transactions on Communications}, vol.~63, no.~5, pp. 1798--1810, May 2015.

\bibitem{7769255}
J.~Qian, F.~Gao, G.~Wang, S.~Jin, and H.~Zhu, ``Noncoherent detections for
  ambient backscatter system,'' \emph{IEEE Transactions on Wireless
  Communications}, vol.~16, no.~3, pp. 1412--1422, march 2017.

\bibitem{8532293}
J.~K. Devineni and H.~S. Dhillon, ``Ambient backscatter systems: Exact average
  bit error rate under fading channels,'' \emph{IEEE Transactions on Green
  Communications and Networking}, vol.~3, no.~1, pp. 11--25, March 2019.

\bibitem{8721108}
J.~Qian, Y.~Zhu, C.~He, F.~Gao, and S.~Jin, ``Achievable rate and capacity
  analysis for ambient backscatter communications,'' \emph{IEEE Transactions on
  Communications}, vol.~67, no.~9, pp. 6299--6310, Sept. 2019.

\bibitem{loraphysical2018}
G.~Ferré and A.~Giremus, ``{LoRa} physical layer principle and performance
  analysis,'' in \emph{2018 25th IEEE International Conference on Electronics,
  Circuits and Systems (ICECS)}, Dec. 2018, pp. 65--68.

\bibitem{proakis2007digital}
J.~G. Proakis and M.~Salehi, \emph{Digital Communications}, 5th~ed.\hskip 1em
  plus 0.5em minus 0.4em\relax New York, NY, USA: McGraw-Hill, 2007.

\bibitem{berLoRaExactSensor}
C.~Ferreira~Dias, E.~Rodrigues~de Lima, and G.~Fraidenraich, ``Bit error rate
  closed-form expressions for {LoRa} systems under {Nakagami} and {Rice} fading
  channels,'' \emph{Sensors}, vol.~19, no.~20, p. 4412, Oct. 2019.

\bibitem{asymptoticBERLoRa2021}
V.~Savaux and G.~Ferré, ``Simple asymptotic {BER} expressions for {LoRa}
  system over {Rice} and {Rayleigh} channels,'' in \emph{2021 Wireless
  Telecommunications Symposium (WTS)}, Apr. 2021, pp. 1--4.

\bibitem{7closedform}
T.~Elshabrawy and J.~Robert, ``Closed-form approximation of {LoRa} modulation
  {BER} performance,'' \emph{IEEE Communications Letters}, vol.~22, no.~9, pp.
  1778--1781, Sept. 2018.

\bibitem{computablelora2021}
J.~Courjault, B.~Vrigneau, O.~Berder, and M.~R. Bhatnagar, ``A computable form
  for {LoRa} performance estimation: Application to {Ricean} and {Nakagami}
  fading,'' \emph{IEEE Access}, vol.~9, pp. 81\,601--81\,611, Apr. 2021.

\bibitem{baruffa2020error}
G.~Baruffa, L.~Rugini, L.~Germani, and F.~Frescura, ``Error probability
  performance of chirp modulation in uncoded and coded lora systems,''
  \emph{Digital Signal Processing}, vol. 106, p. 102828, Nov. 2020.

\bibitem{9693529}
S.~An, H.~Wang, Y.~Sun, Z.~Lu, and Q.~Yu, ``Time domain multiplexed lora
  modulation waveform design for iot communication,'' \emph{IEEE Communications
  Letters}, vol.~26, no.~4, pp. 838--842, April 2022.

\bibitem{loraTheoreticalAndExperimental}
M.~J. Faber, K.~M. van~der Zwaag, W.~G.~V. dos Santos, H.~R. d.~O. Rocha,
  M.~E.~V. Segatto, and J.~A.~L. Silva, ``A theoretical and experimental
  evaluation on the performance of {LoRa} technology,'' \emph{IEEE Sensors
  Journal}, vol.~20, no.~16, pp. 9480--9489, Aug. 2020.

\bibitem{lorainterference2019}
O.~Afisiadis, M.~Cotting, A.~Burg, and A.~Balatsoukas-Stimming, ``On the error
  rate of the {LoRa} modulation with interference,'' \emph{IEEE Transactions on
  Wireless Communications}, vol.~19, no.~2, pp. 1292--1304, Feb. 2020.

\bibitem{loraSameSFInterference2018}
T.~Elshabrawy and J.~Robert, ``Analysis of {BER} and coverage performance of
  {LoRa} modulation under same spreading factor interference,'' in \emph{2018
  IEEE 29th Annual International Symposium on Personal, Indoor and Mobile Radio
  Communications (PIMRC)}, Sept. 2018, pp. 1--6.

\bibitem{loraImperfectOrtho2018}
D.~Croce, M.~Gucciardo, S.~Mangione, G.~Santaromita, and I.~Tinnirello,
  ``Impact of {LoRa} imperfect orthogonality: Analysis of link-level
  performance,'' \emph{IEEE Communications Letters}, vol.~22, no.~4, pp.
  796--799, Apr. 2018.

\bibitem{Benkhalifa:22}
F.~Benkhelifa, Y.~Bouazizi, and J.~A. McCann, ``How orthogonal is {LoRa}
  modulation?'' \emph{IEEE Internet of Things Journal}, vol.~9, no.~20, pp.
  19\,928--19\,944, May 2022.

\bibitem{9641890}
A.~S. Ali, L.~Bariah, S.~Muhaidat, P.~Sofotasios, and M.~Al-Qutayri,
  ``Performance analysis of {LoRa}-backscatter communication system in awgn
  channels,'' in \emph{2021 4th International Conference on Advanced
  Communication Technologies and Networking (CommNet)}, Dec. 2021, pp. 1--5.

\bibitem{ETSI}
``Electromagnetic compatibility and radio spectrum matters ({ERM}); shortrange
  devices ({SRD}); radio equipment to be used in the 25 mhz to1000 mhz
  frequency range with power levels ranging up to 500 mw;part 1: Technical
  characteristics and test methods,'' Eur. Telecommun.Stand. Inst., Sophia
  Antipolis, France, Standard, 2012.

\bibitem{harris1998handbook}
J.~W. Harris and H.~St{\"o}cker, \emph{Handbook of mathematics and
  computational science}.\hskip 1em plus 0.5em minus 0.4em\relax Springer
  Science \& Business Media, 1998.

\bibitem{8835951}
T.~Elshabrawy, P.~Edward, M.~Ashour, and J.~Robert, ``On the different
  mathematical realizations for the digital synthesis of lora-based
  modulation,'' in \emph{European Wireless 2019; 25th European Wireless
  Conference}, May 2019, pp. 1--6.

\bibitem{digitalCommunication2012}
E.~A. Lee and D.~G. Messerschmitt, \emph{Digital communication}.\hskip 1em plus
  0.5em minus 0.4em\relax Springer Science \& Business Media, 2012.

\bibitem{benedetto1999principles}
S.~Benedetto and E.~Biglieri, \emph{Principles of digital transmission: with
  wireless applications}.\hskip 1em plus 0.5em minus 0.4em\relax Springer
  Science \& Business Media, 1999.

\bibitem{705532}
M.~Simon and M.~Alouini, ``A unified approach to the performance analysis of
  digital communication over generalized fading channels,'' \emph{Proceedings
  of the IEEE}, vol.~86, no.~9, pp. 1860--1877, 1998.

\bibitem{1964handbook}
M.~Abramowitz and I.~A. Stegun, \emph{Handbook of mathematical functions with
  formulas, graphs, and mathematical tables}.\hskip 1em plus 0.5em minus
  0.4em\relax US Government printing office, 1964, vol.~55.

\bibitem{gudbjartsson1995rician}
H.~Gudbjartsson and S.~Patz, ``The {Rician} distribution of noisy {MRI }data,''
  \emph{Magnetic resonance in medicine}, vol.~34, no.~6, pp. 910--914, 1995.

\bibitem{9915518}
A.~Subhash, S.~Kalyani, Y.~H. Al-Badarneh, and M.-S. Alouini, ``On the
  asymptotic performance analysis of the k-th best link selection over
  non-identical non-central chi-square fading channels,'' \emph{IEEE
  Transactions on Communications}, vol.~70, no.~11, pp. 7191--7206, Nov. 2022.

\bibitem{nNakagami2007n}
G.~K. Karagiannidis, N.~C. Sagias, and P.~T. Mathiopoulos,
  ``{N}$\ast${Nakagami}: A novel stochastic model for cascaded fading
  channels,'' \emph{IEEE Transactions on Communications}, vol.~55, no.~8, pp.
  1453--1458, Aug. 2007.

\bibitem{tableofintegral}
I.~S. Gradshteyn and I.~M. Ryzhik, \emph{Table of integrals, series, and
  products}.\hskip 1em plus 0.5em minus 0.4em\relax Academic press, 2014.

\bibitem{rabinowitz1959tables}
P.~Rabinowitz and G.~Weiss, ``Tables of abscissas and weights for numerical
  evaluation of integrals of the form$\int_0^\infty
  e^{-x}x^nf\of{x}\mathrm{d}x$,'' \emph{Mathematical Tables and Other Aids to
  Computation}, pp. 285--294, 1959.

\bibitem{besselkbound2017}
Z.-H. Yang and S.-Z. Zheng, ``The monotonicity and convexity for the ratios of
  modified {Bessel} functions of the second kind and applications,''
  \emph{Proceedings of the American Mathematical Society}, vol. 145, no.~7, pp.
  2943--2958, Jan. 2017.

\bibitem{goldsmith2005wireless}
A.~Goldsmith, \emph{Wireless communications}.\hskip 1em plus 0.5em minus
  0.4em\relax Cambridge university press, 2005.

\bibitem{digitalcommunicationoverfadingchannell}
M.~K. Simon and M.-S. Alouini, \emph{Digital communication over fading
  channels}.\hskip 1em plus 0.5em minus 0.4em\relax John Wiley \& Sons, 2005,
  vol.~95.

\end{thebibliography}

\end{document}